\documentclass[useAMS,usenatbib]{mn2e}
\usepackage{graphicx} 
\usepackage{dcolumn}  
\usepackage{bm}       
\usepackage{amssymb,amsmath}  
\title[Tracing The Sound Horizon Scale With Photometric Redshift Surveys]{Tracing The Sound Horizon Scale With Photometric Redshift Surveys}
\author[E. S\'anchez et al.]
{E. S\'anchez$^1$\thanks{E-mail:eusebio.sanchez@ciemat.es}, 
 A. Carnero$^1$, J.~Garc\'{\i}a-Bellido$^{2,3}$, E. Gazta\~naga$^4$, F. de Simoni$^5$,
 \newauthor 
 M. Crocce$^4$, A. Cabr\'e$^6$, P. Fosalba$^4$, D. Alonso$^2$ \\
 $^1$Centro de Investigaciones En\'ergeticas, Medioambientales y Tecnol\'ogicas (CIEMAT), Madrid, Spain\\
 $^2$Instituto de F\'{\i}sica Te\'orica (UAM-CSIC), Madrid, Spain \\
 $^3$D\'epartement de Physique Th\'eorique, Universit\'e de Gen\`eve, 24 quai Ernest Ansermet, CH--1211 Gen\`eve 4, Switzerland\\
 $^4$Institut de Ci\`encies de l'Espai (IEEC-CSIC), Barcelona, Spain \\
 $^5$Observat\'orio Nacional, Rio de Janeiro, Brazil \\
 $^6$University of Pennsylvania, Philadelphia, USA\\
}
\begin{document}

\date{\today}
\pagerange{1--12} \pubyear{2002}
\maketitle

\begin{abstract}
We propose a new method for cosmological parameters extraction using 
the baryon acoustic oscillation scale as a standard ruler in deep 
galaxy surveys with photometric determination of redshifts. The method 
consists in a simple empirical parametric fit to the angular 2-point 
correlation function $\omega(\theta)$. It is parametrized as a power law 
to describe the continuum plus a Gaussian to describe the BAO bump. The
location of the Gaussian is used as the basis for the measurement of the
sound horizon scale. This method, although simple, actually provides a 
robust estimation, since the inclusion of the power law and the use of
the Gaussian removes the shifts which affect the local maximum. We discuss 
the effects of projection bias, non-linearities, redshift space distortions 
and photo-z precision, and apply our method to a mock catalog of 
the Dark Energy Survey, built upon a large N-body simulation provided
by the MICE collaboration. We 
discuss the main systematic errors associated to our method and show that 
they are dominated by the photo-z uncertainty.
\end{abstract}

\section{Introduction}
\label{intro}
Primordial perturbations generated acoustic waves in the photon-baryon fluid 
until decoupling $(z\sim1100)$. At this time the photons decouple from the 
baryons creating a high density region from the original source of 
perturbation, at a distance given by the sound horizon length. This high 
density profile shows as a peak associated to the sound horizon scale 
in the galaxies 
spatial two-point statistics in the configuration space, and as a series of 
oscillations in Fourier space, which can be used as a cosmological standard 
ruler. Recently, this feature has been applied to constrain cosmological 
parameters using several methods \citep{2005ApJ...633..560E,
2006A&A...449..891H,2006A&A...459..375H,2007MNRAS.381.1053P,
2007MNRAS.378..852P,2008ApJ...676..889O,2009MNRAS.399.1663G,
2009MNRAS.400.1643S}. Those observations together with WMAP7 and Type Ia 
supernovae yield to the concordance cosmology: a spatially flat and late 
time accelerated universe \citep{2010arXiv1001.4538K}.

In order to achieve high accuracy on the cosmological parameters,
several galaxy surveys will be done estimating redshifts from
photometric data, like 
DES \citep{2005astro.ph.10346T}, PanSTARRS \citep{2000PASP..112..768K}, LSST 
\citep{2003NuPhS.124...21T} or PAU \citep{Benitez:2008fs}. This higher 
accuracy will be possible 
thanks to a larger volume and number of observed galaxies compared to previous
spectroscopic surveys such as SDSS \citep{2000AJ....120.1579Y} or 
2dF \citep{2001MNRAS.328.1039C}, even if these photometric redshifts 
(photo-z) will have lower precision compared to their spectroscopic 
counterparts. The 
photo-z error depends mostly on the range of wavelengths covered by the 
filters in which the observations are done, the number of filters and 
their photometric error. As 
an example, the CFHTLS survey found, in the range $0.2 \leq z \leq 1.5$, a 
redshift error $\sigma_{z} \sim 0.03(1+z)$. This error was obtained with 
observations in the 
optical and the near infrared \citep{2006A&A...457..841I}. An error of this 
magnitude represents an uncertainty in
the radial position of the galaxies which makes it impossible to infer the
true 3-dimensional clustering pattern. Therefore, the analysis of angular 
statistics, like the two-point angular correlation 
function $\omega(\theta)$ and the angular power spectrum $C_{\ell}$,
is required for these surveys.

In this paper, a new method based on an empirical parametrization of
$\omega(\theta)$  in redshift shells is proposed. The goal is to recover
the angle corresponding to the BAO scale as a function of the redshift, and
obtain the properties of the dark energy from its evolution. 

The method is designed to be used as a standard ruler only, {\it i.e.},  we 
do not try
to use the whole shape of the correlation function or the power spectrum. The 
reason is that although the strict standard ruler is less sensitive to the
cosmological parameters, it seems also more robust against systematic 
uncertainties, since the only observation to be used is the position of the
BAO peak in the two-point angular correlation function, and nothing else. It 
is unclear which measurement will be more sensitive at the end of the 
day, including all the systematic errors, whether the full description or the 
standard ruler~\citep{Rassat:2008ja,2008MNRAS.390.1470S}.

The main difficulty of an approach using the position of the BAO
peak as a standard ruler is the extent at which the ruler remains
``standard''. The position of the BAO peak is subject to several
effects that shift its location and reduce its contrast. The most 
important among these is the projection offset due to the redshift bin 
width used to slice the data. This ultimately changes the $\omega(\theta)$
shape for wide enough binning leaving a shoulder shape with no local
maximum.

Our parametric approach is nonetheless able to measure the BAO scale even for
wide redshift bins, and correct for the shift due to the projection 
effect, recovering the input cosmology. The proposed method has been 
tested against theoretical calculations and also applied to mock surveys 
from N-body simulations. Nevertheless, it should be useful to study BAO 
as a standard ruler in any photometric redshift survey.

\section{Angular Clustering}
\label{sec:theoryAngCor}
In complete analogy to the spatial correlation function $\xi(r)$, the 
angular correlation function $\omega(\theta)$ is defined as the excess 
joint probability that two point sources (\textit{e.g.} galaxies) are 
found in two solid angle elements $d\Omega_1$ and $d\Omega_2$ with 
angular separation $\theta$ compared to a homogeneous Poisson 
distribution \citep{1980lssu.book.....P}. From this definition, it 
is easy to find the relation between $\xi(r)$ and $\omega(\theta)$,

\begin{equation}
 \omega(\theta) = \int_{0}^{\infty}\!\! dz_1\, \phi(z_1) \int_{0}^{\infty}\!\! dz_2\, \phi(z_2)\, \xi(r; \bar{z}) \,\, ,
\label{eq:xir2wtheta}
\end{equation}

\noindent
where $\phi(z)$ is the redshift shell selection function normalized to 
unity and $\bar{z} = (z_1+z_2)/2$. This relation assumes the time evolution 
of the real-space spatial correlation 
function is small within the redshift shell under 
analysis, \textit{i.e.}, $\xi(r;z_1) \approx \xi(r;z_2)$ (for a relation 
without approximations see \citealt{2004ApJ...606....1M}). The comoving 
distance between two point sources for a spatially flat cosmology 
is (for a generalization to other spatial curvatures see 
again \citealt{2004ApJ...606....1M})

\begin{equation}
r = \sqrt{\chi(z_1)^{2} + \chi(z_2)^2 - 2\chi(z_1)\chi(z_2)\cos\theta} \,\, .
\label{dist1}
\end{equation}

The comoving radial distance $\chi(z)$, assuming a constant dark energy 
equation of state parameter $w$~\footnote{Some of the cosmological models used 
in this analysis are non-flat and others have non-constant $w$. In these cases, the
full calculation has been used.}, is defined as

\begin{equation}
\chi(z) = \frac{c}{H_0} \int_{0}^{z}\frac{dz}{\sqrt{\Omega_m (1+z)^3 + \Omega_{\Lambda}(1+z)^{3(1+w)}}} \,\, .
\label{comovingRadial}
\end{equation}

\noindent
As usual, $c$ is the speed of light, $H_0$ is the present time Hubble 
parameter, $\Omega_{m}$ is the matter density parameter and 
$\Omega_{\Lambda} = (1-\Omega_{m})$ is the dark energy density 
parameter. With the relation between $\omega(\theta)$ and $\xi(r)$ in 
hand, one needs a model for the two quantities that appear on 
Eq. (\ref{eq:xir2wtheta}): the spatial correlation function, including
the effects of non-linearities and the selection function 
$\phi(z)$ that takes into account observational effects.

The spatial correlation funcion is given by,

\begin{equation}
\xi(r;z) = \int_{0}^{\infty}\!\! \frac{dk}{2\pi^{2}} k^{2}j_{0}(kr)P(k;z)\,\, ,
\label{Power2CorFourier}
\end{equation}

\noindent
where $j_{0}(x)$ is the spherical Bessel function of zeroth order and 
$P(k)$ the power spectrum. 

In linear theory, valid at large distances, 
$P_{L}(k)=b^2D^2(z) P(k,z=0)$, with $b$ a constant bias factor and 
$D(z)=\delta(z)/\delta(z=0)$, the growth factor of the
density contrast, $\delta(z)$. Since we are only concerned with the BAO peak
position and not the clustering amplitude, the bias $b$ between
luminous and dark matter will be set to unity throughout our 
analysis. We are assuming, then, that it does not evolve much in a redshift 
shell and that it is scale independent, see section~\ref{subsub:other}, as well as 
neglecting peak shifts in the 3-d correlation of galaxies/halos with respect 
to that of dark matter in front of projection effects 
\citep{2008PhRvD..77d3525S,2008MNRAS.390.1470S}. This is a very good
approximation for the scales we are dealing with, as has been measured in
~\citet{2009MNRAS.392..682C}. A brief discussion of the systematic error that
is induced by this assumption can be found in section~\ref{subsec:sys}.

We introduce nonlinear matter clustering only through a damping of the
BAO wiggles in the linear spectrum, i.e., 
$P_{L}\rightarrow P_{L} {\rm e}^{-k^2 \sigma^2_v(z)/2}$ 
 \citep{1996ApJ...472....1B,2008PhRvD..77b3533C,2008PhRvD..77f3530M,
2007ApJ...664..660E,2006PhRvD..73f3519C}. In doing this 
we discard the contribution 
to $\omega(\theta)$ of the additive mode-coupling term to $P(k)$, discussed 
in \citet{2008PhRvD..77b3533C} (see also \citealt{2009MNRAS.400.1643S}), what is 
justified because it has a negligible impact in the {\it angular} position of 
the BAO peak in particular in comparison to projection 
effects \citep{crocce10}. The quantity $\sigma_v$ characterizes the scale 
where non-linear effects become significant. It is related to the linear 
power spectrum via 

\begin{equation}
 \sigma_v(z) = \left[ \frac{1}{6\pi^{2}} \int_{0}^{\infty}\!\! 
dk P_{L}(k;z) \right]^{-1/2} \,\, .
\end{equation}

In turn, the covariance matrix of $\omega(\theta)$,
${\rm Cov}_{\theta \theta^{\prime}} \equiv \langle
\omega(\theta)\omega(\theta^{\prime}) \rangle,$ for a given survey can be
estimated by ({\it e.g.}~\citealt{Cabre:2007rv},~\citealt{crocce10}),

\begin{eqnarray}
\nonumber
\lefteqn{{\rm Cov}_{\theta \theta^{\prime}} =} \\ 
\lefteqn{\sum_{l \ge 0} \frac{2 (2l+1) P_l(\cos(\theta)) P_l(\cos(\theta^{\prime}))}{(4\pi)^{2} f_{sky}} \, \,\left[C(l)+\frac{1}{N/\Delta \Omega}\right]^2}
\label{eq:CovW}
\end{eqnarray}

\noindent
where $f_{sky}$ is the fraction of the sky covered
by the survey and the ratio $N/\Delta \Omega$ is the number of galaxies
per unit of solid angle in units of srad$^{-1}$. What we call statistical error along this 
analysis is computed using this expression and applied to 
$\omega(\theta)$. A detailed analysis of these expressions and
their applicability is given in \citet{crocce10}.

\section{The Sound Horizon Scale}
\label{sec:soundhorizonscale}
The strength of the standard ruler method lays in the potential to relate
rather straightforwardly the acoustic peak position in the correlation 
function of galaxies to the sound horizon scale at decoupling. However, these 
two quantities are not exactly equal~\citep{2008MNRAS.390.1470S,
2009PhRvD..80f3508P,2009arXiv0910.5005S,2008ApJ...686...13S,
2007ApJ...665...14S}. Thus, in 
any standard ruler analysis it is crucial to distinguish between the 
following two angular scales, $\theta_{BAO}\equiv r_{s}/\chi(z)$, the 
angular scale corresponding to the sound horizon at the drag epoch and
$\theta_{p}$, the location of the BAO peak in the angular correlation 
function, defined as the local maximum around the expected angular 
scale. The peak position in the linear angular
correlation function $\theta_{p}$ only approaches the sound horizon
scale $\theta_{BAO}$ for infinitesimal shells, but a residual difference
which ranges from 1 to 2 \% is found~\citep{2008MNRAS.390.1470S}, depending
on the cosmological parameters. 

Moreover, if the survey under
analysis is performed with photo-z, there will be no such thing as an
infinitesimal shell, due to the uncertainty in redshift associated to the
photo-z. On the other hand, the impact of the redshift space distortions on 
$\omega(\theta)$ is large even for large angular scales, as has been pointed
out by~\citet{Nock:2010gr}, contributing also to these effects.

\begin{figure}
\centering
\includegraphics[width=0.50\textwidth]{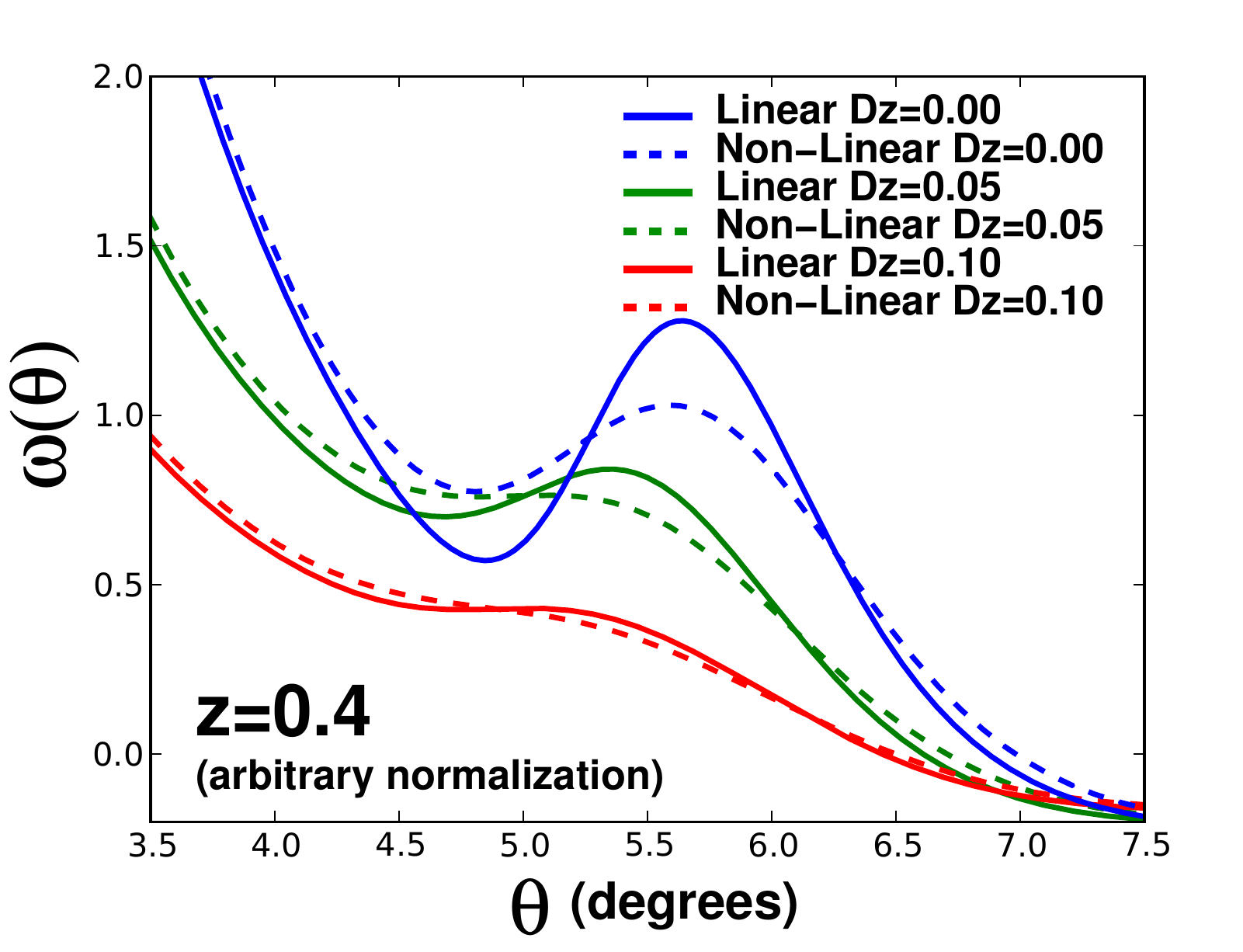}
\caption{The impact of non-linearities and projection effects on the
  local maximum of the angular correlation function. The continuous
  line is the linear prediction and the dashed line the non-linear
  model from section~\ref{sec:theoryAngCor}. The figure assumes a redshift bin
  centered at $z=0.4$ with the upper lines (blue) correspoding to an infinitesimal
  shell, the middle lines (green) to a shell
  width of $\Delta z=0.05$ and the lower ones (red) to $\Delta z=0.1$. Two effects 
  are seen. First, the position of the peak moves
  towards lower angles when $\Delta z$ grows. Second, its amplitude
  reduces until the local maximum disappears, but a plateau survives
  instead. The dominant effect is the projection, while the impact of the
  non-linearities is much smaller.\label{fig:nonlinear}}
\end{figure}

A large photo-z error (\textit{e.g.} $\sigma_{z}\geq0.03$), will require a 
wide redshift shell. But it is well known that this fact induces
two effects, which are shown in Figure~\ref{fig:nonlinear}. The first one 
is that the peak position shifts towards
smaller angles, and the displacement is larger for larger widths. Second, the 
amplitude of the peak gets reduced until the local maximum 
disappears~\citep{2008PhRvD..77b3512L,2009PhRvD..79f3508S}, and only a 
shoulder in $\omega(\theta)$ remains. As an order of magnitude, we have
computed that for $\sigma_{z}=0.01 (1+z)$ 
there will be a well defined local maximum from mean redshift 
$z \gtrsim 0.6$ and  in the case $\sigma_{z}=0.03 (1+z)$ it will be there
only for $z\gtrsim 1.3$. 

These results suggest we should measure $\theta_{BAO}$  
using an alternative method to the position of the local maximum 
of $\omega(\theta)$.  Currently, the ratio of the power spectrum 
to a smooth fit is used, to avoid the effect of the tilting function, which 
changes the peak position~\citep{2008ApJ...686...13S,2009arXiv0910.5005S,
2008PhRvD..77b3533C,2008PhRvD..77d3525S}. Here we present an alternative 
method based on the angular correlation function. This 
will be the subject of the subsequent sections.

\section{Method to Recover $\btheta_{\bf BAO}$}
\label{sec:method}
Trying to overcome the problems induced by the photo-z, we propose a 
new method based on an empirical parametrization of $\omega(\theta)$, which can 
be used even when there is no local maximum. The most important points of
the method, the parametrization of $\omega(\theta)$ and the correction for the
projection effect due to the width of the redshift bin are explained in
detail in sections~\ref{subsec:parametrization} 
and~\ref{subsec:correction}, respectively. The full recipe to obtain the 
BAO scale as a function of redshift is as follows:

\begin{enumerate}
\item Divide the full galaxy sample in redshift bins.

\item Compute the angular two-point correlation function in each redshift
bin. 

\item Parametrize the correlation function using the expression:
\begin{equation}
\label{eq:param}
\omega(\theta) = A + B \theta^{\gamma} + C e^{-(\theta - \theta_{FIT})^{2}/2\sigma^{2}}
\end{equation}

\noindent
and perform a fit to $\omega(\theta)$ with free parameters 
$A$, $B$, $C$, $\gamma$, $\theta_{FIT}$ and $\sigma$. 

\item The BAO scale is estimated using the parameter $\theta_{FIT}$, and 
correcting it for the projection effect (see Eq.~\ref{eq:corrtheta} below). The 
BAO scale as a function of the redshift is the only parameter needed
to apply the standard ruler method. The cosmological interpretation of the
other parameters is limited, since this is an empirical description, valid 
only in a neighborhood of the BAO peak, as is described in 
section~\ref{subsec:parametrization}.

\item Fit cosmological parameters to the evolution of the corrected
$\theta_{BAO}$ with $z$.
\end{enumerate}

  \subsection{Parametrization of $\bomega(\btheta)$}
  \label{subsec:parametrization}

\begin{figure*}
\centering
\leavevmode
\begin{tabular}{ccc}
\includegraphics[width=0.31\textwidth]{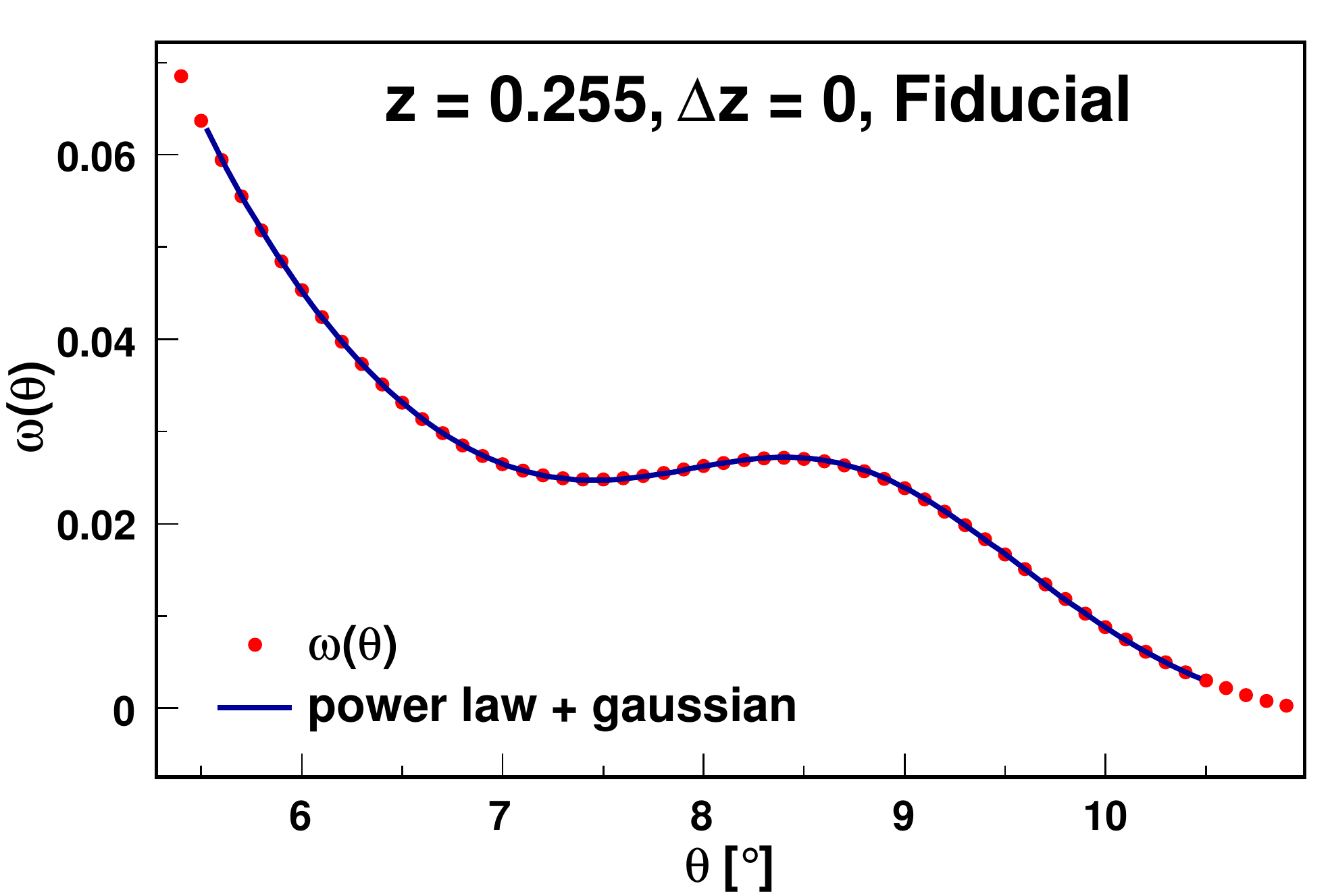} &
\includegraphics[width=0.31\textwidth]{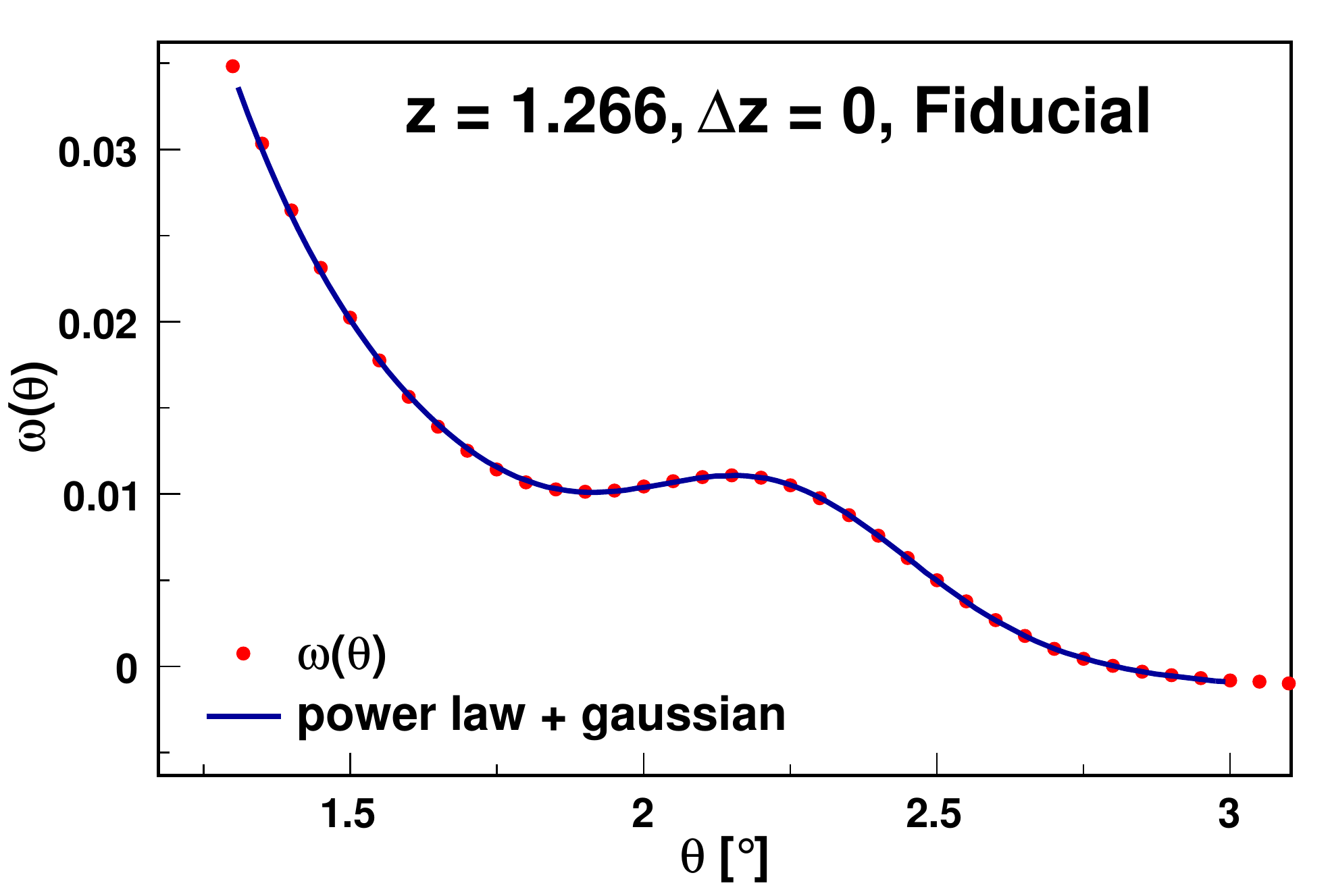} &
\includegraphics[width=0.31\textwidth]{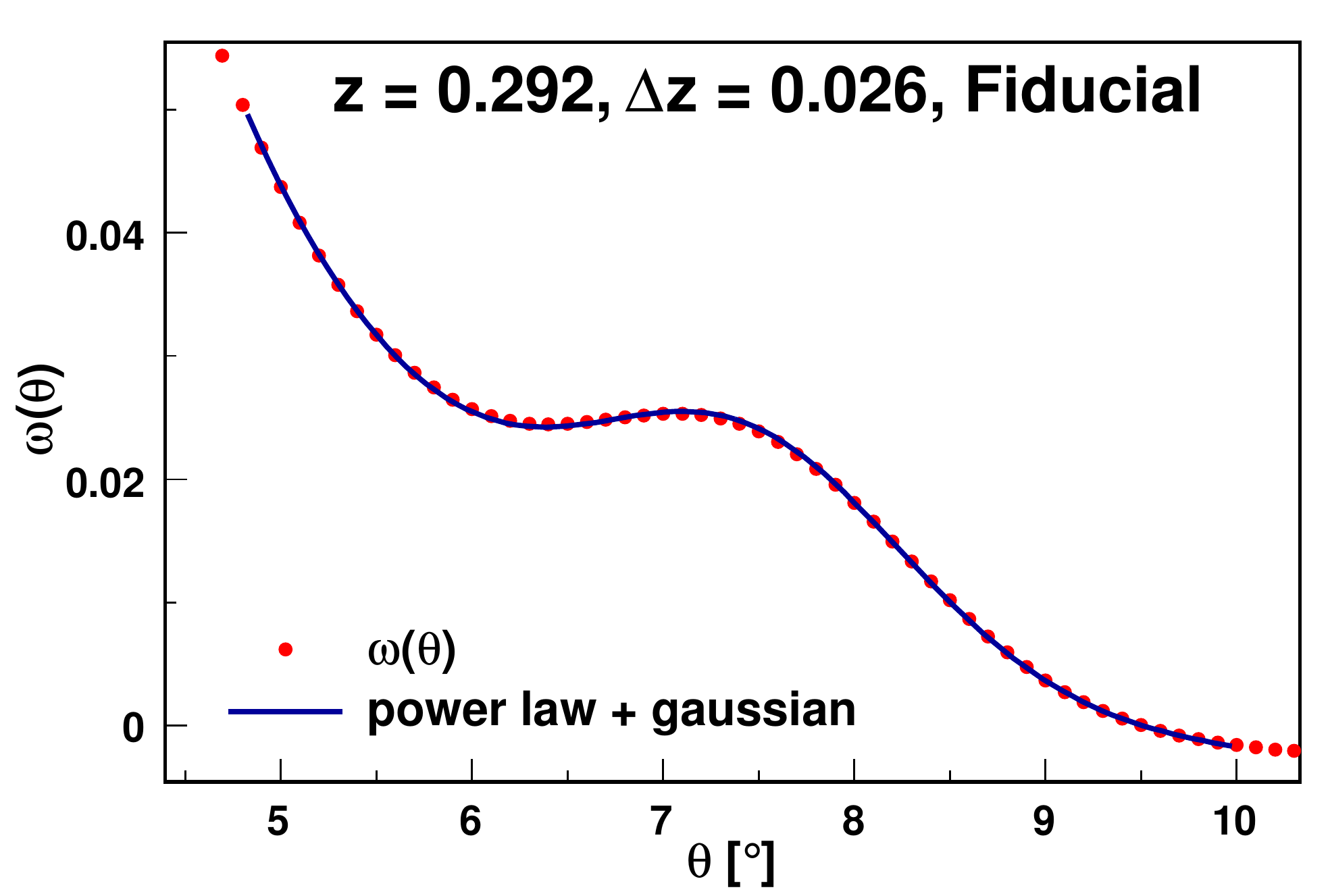} \\ 
\includegraphics[width=0.31\textwidth]{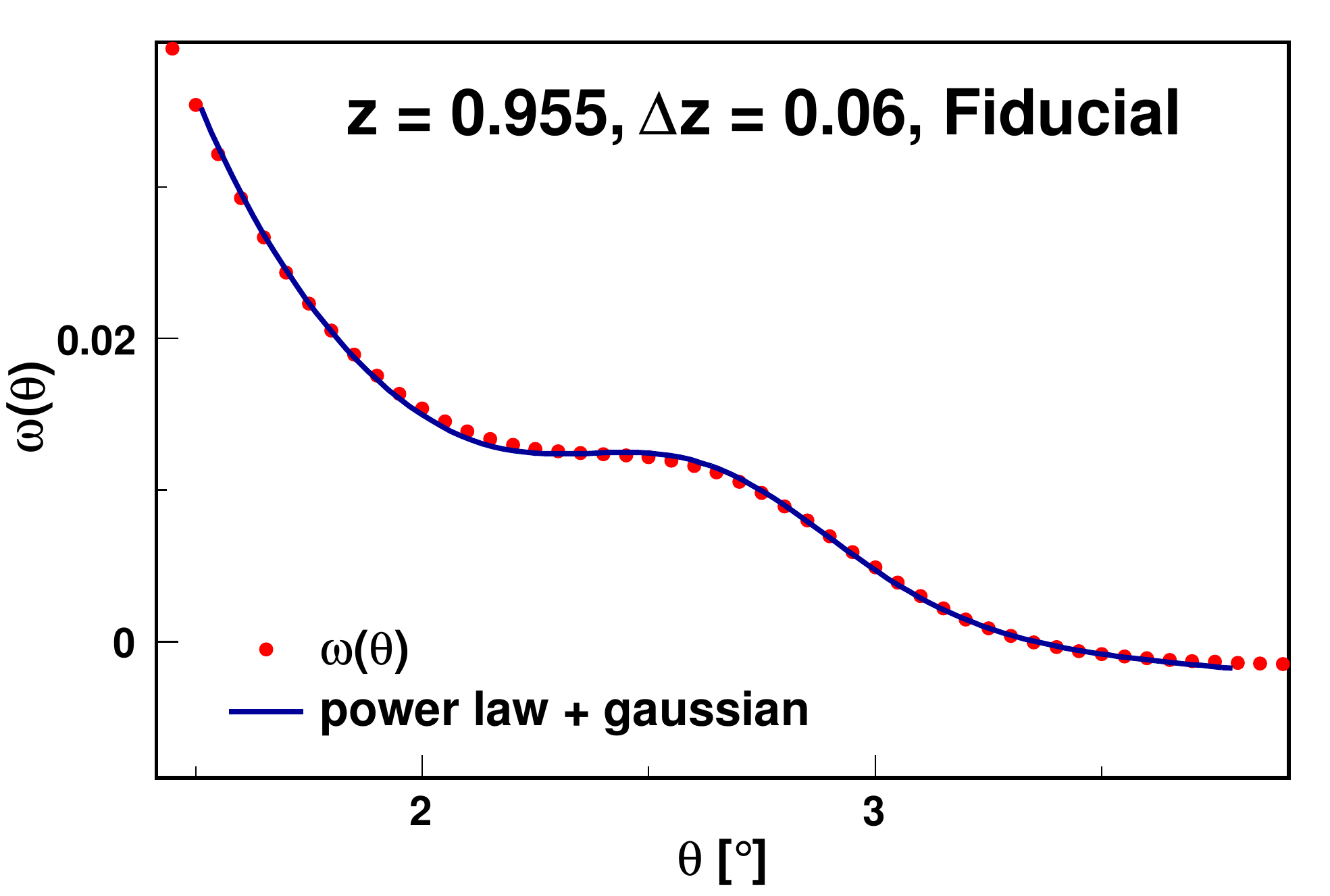} &
\includegraphics[width=0.31\textwidth]{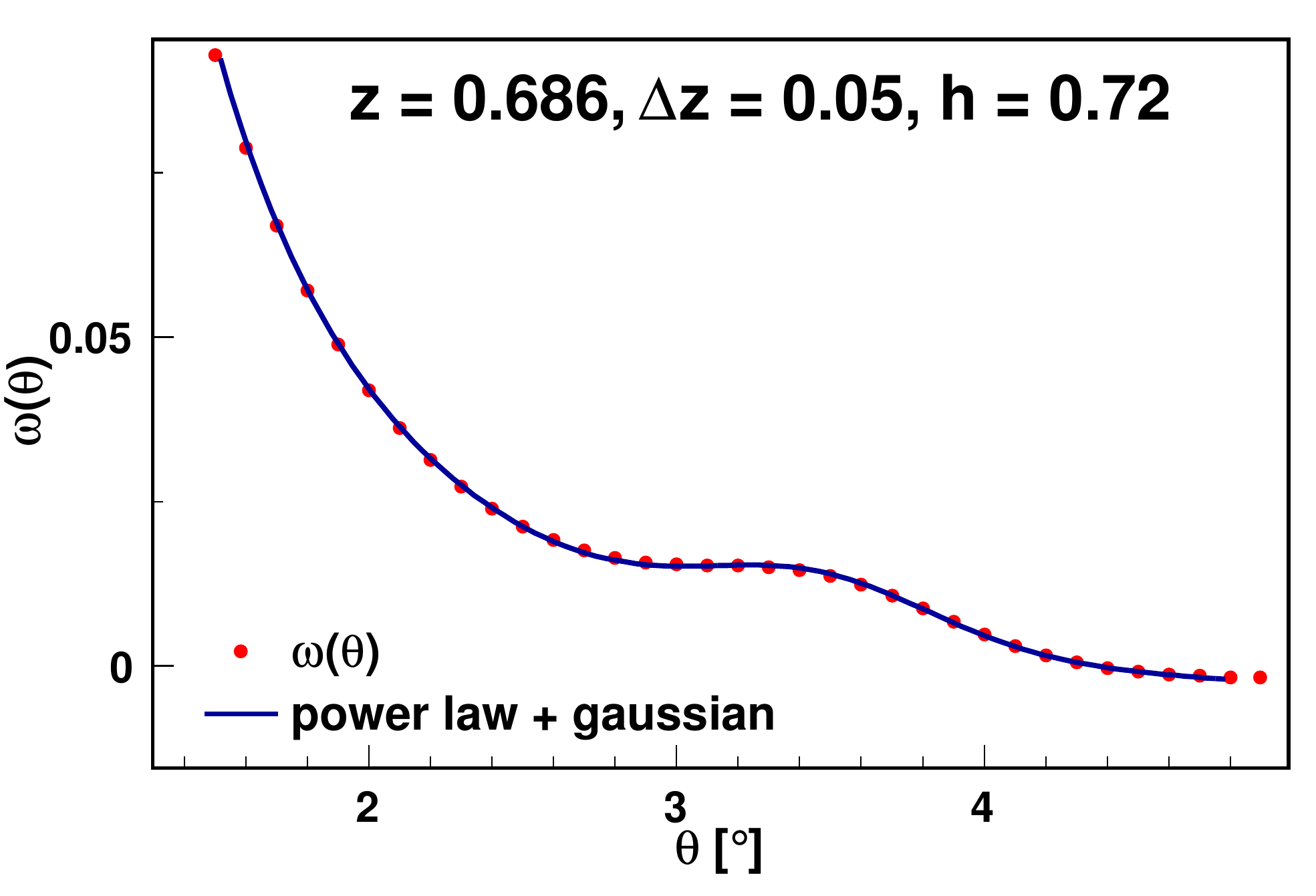} &
\includegraphics[width=0.31\textwidth]{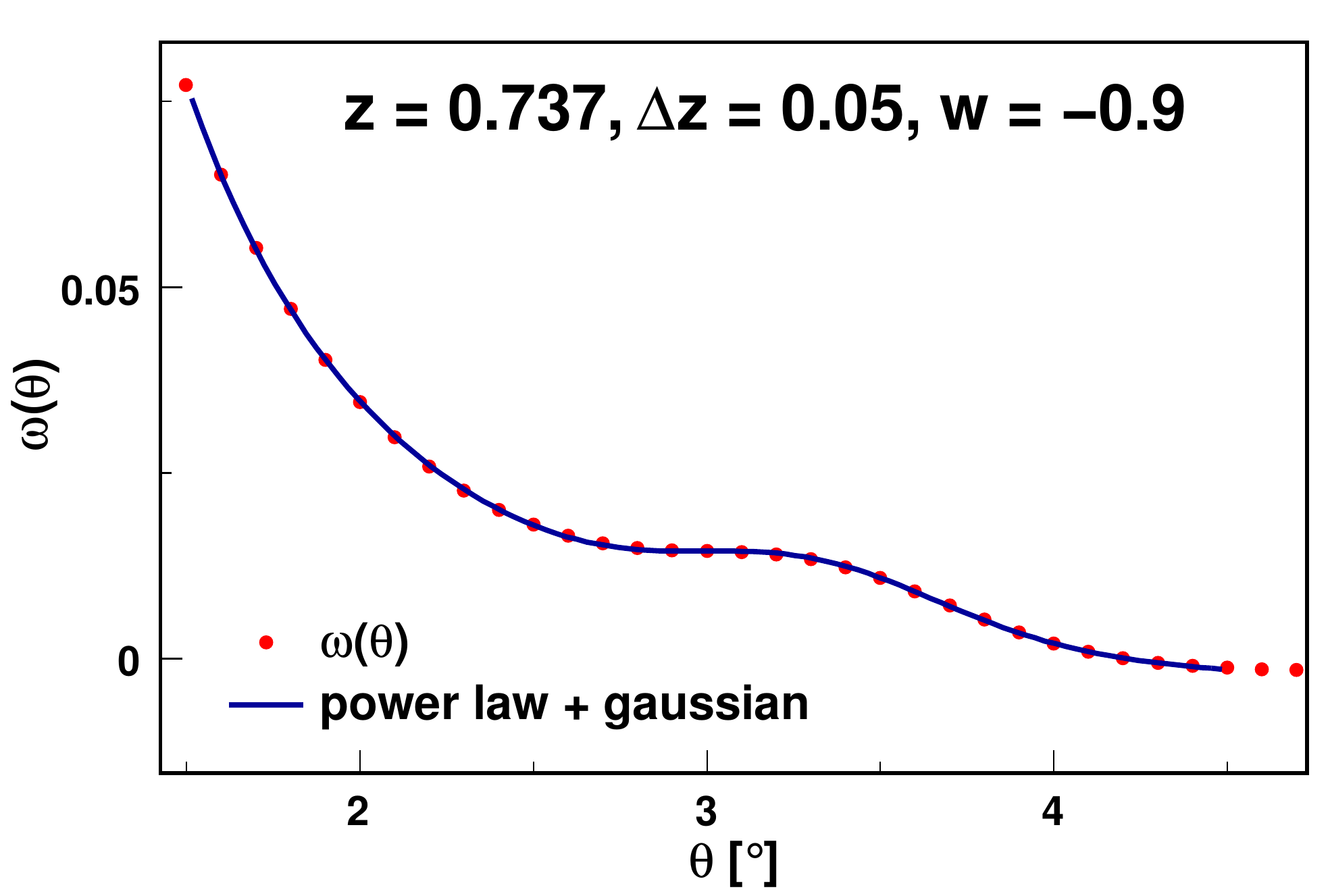} \\
\end{tabular}
\caption{Goodness of the parametrization for several redshifts, several
bin widths and several cosmologies.\label{fig:param} }
\end{figure*}

The correlation function $\omega(\theta)$ is parametrized as in 
Eq. (\ref{eq:param}). A power law is used to describe the shape of the 
correlation function
before and after the peak, and a Gaussian to describe the BAO feature. The
mean of the Gaussian is used to locate the BAO scale. The parameter $A$
takes into account that the correlation function can go to negative values
after the BAO peak. The parameters $B$ and $C$ describe the relative weight of
the two terms of the parametrization. The parameter $\gamma$ is the index
of the power law, while $\theta_{FIT}$ and $\sigma$ are the mean and the
width of the Gaussian, respectively.

\begin{table}
\centering
\begin{tabular}{ccccccc}
\hline
$h$ & $\Omega_{M}$ & $\Omega_{b}$ & $\Omega_{k}$ & $w_{0}$ & $w_{a}$ & $n_s$ \\
\hline\hline
0.70 & 0.25 & 0.044 &  0.00   & $-$1.00 &  0.0   & 0.95 \\
0.68 & ~    & ~     & ~       & ~       & ~      & ~    \\
0.72 & ~    & ~     & ~       & ~       & ~      & ~    \\
~    & 0.20 & ~     & ~       & ~       & ~      & ~    \\ 
~    & 0.30 & ~     & ~       & ~       & ~      & ~    \\ 
~    & ~    & 0.040 & ~       & ~       & ~      & ~    \\
~    & ~    & 0.048 & ~       & ~       & ~      & ~    \\
~    & ~    & ~     & +0.01   & ~       & ~      & ~    \\
~    & ~    & ~     & $-$0.01 & ~       & ~      & ~    \\
~    & ~    & ~     & ~       & $-$0.90 & ~      & ~    \\
~    & ~    & ~     & ~       & $-$1.10 & ~      & ~    \\
~    & ~    & ~     & ~       & ~       & $-$0.1 & ~    \\
~    & ~    & ~     & ~       & ~       & +0.1   & ~    \\
~    & ~    & ~     & ~       & ~       & ~      & 1.00 \\
\hline
\end{tabular}
\caption{Summary of the 14 cosmological models used to test the 
method.  Where empty, the fiducial values (first line)
are assumed.\label{tab:cosmomodels}}
\end{table}

We have tested the goodness of this parametrization in a redshift 
interval which ranges from 0.2 to 1.4, for a wide range of widths of 
the redshift bins (from 0 to 0.2) and for 14 cosmological models, which 
are listed in Table~\ref{tab:cosmomodels}. The  parametrization is good
within 1\%, {\it i.e.}, the fit is good (the value of $\chi^2$/ndof 
ranges from 0.98 to 1.01, and the probabilities of the fit from
0.6 to 0.9) when the error in each 
point of the correlation function is $\sim 1\%$ for all bin
widths and cosmological models. This precision is much better 
than that expected in any realistic redshift survey, since it is much
smaller than the cosmic variance. Several examples of the 
parametrization  can be seen in Fig.~\ref{fig:param}. Other functional
forms have been tested, but all of them are more complex and show
no improvements on the description of the data.

There is a region of stability to choose the 
starting and end points of the fit, which will be described in more detail 
in section~\ref{subsub:sys_method}. This region cannot be made
arbitrarily large, since the description cannot be good for very low 
angles, due to non-linearities and changes in the power law index; neither 
for very large angles, since the correlation function changes its
slope after the BAO peak.

  \subsection{Correcting for Projection Effects}
  \label{subsec:correction}
The projection offset discussed in Sec.~\ref{sec:soundhorizonscale} is 
also present for the parametrization method. If we want to extract cosmology 
from this measurement, it is necessary to
correct the fitted $\theta_{FIT}$ for the projection effect to recover
$\theta_{BAO}$. 

\begin{equation}
\label{eq:corrtheta}
\theta_{BAO}=\alpha \,\, \theta_{FIT},
\end{equation}

\noindent
where $\alpha$ could in principle be a function of redshift, bin width
and cosmology. However, if we want to
do this correction in an unbiased way, it must be cosmology 
independent. We have tested the method for the 14 different cosmologies 
summarized in Table~\ref{tab:cosmomodels}, and for different redshift
bin widths. 

For each of the 14 considered models, each of the considered
redshifts and each of the considered bin widths, we compute the angular 
correlation function. We use Eq.~\ref{eq:xir2wtheta} and include 
non-linearities (see section~\ref{sec:theoryAngCor}). The
calculation includes also the effect of the galaxy distribution with the
redshift, as descibed in section~\ref{sec:simulation}. The redshift 
ranges from 0.2 to 1.4 and the bin width from 0 to 0.2, as has been 
previously stated. We then apply the 
fitting method to obtain the position of the BAO peak. Two main results
are obtained:

\begin{itemize}
\item The shift of $\theta_{FIT}$ with respect to $\theta_{BAO}$ has a
      universal shape, independent of the cosmology. Thus the
      correction function $\alpha$ depends only on redshift and bin
      width, but not on the cosmological 
      model, {\it i.e.}, $\alpha = \alpha(z,\Delta z)$, within a
      precision $< 1\%$.
\item After applying this universal shift, the recovered BAO scale
      agrees with the true value for all cosmologies to a precision of 
      $\leq 0.75\%$. In particular, for infinitesimal bin widths, the
      method is able to compensate the displacement of the BAO bump due to
      the tilting function, correcting the 2\% difference which has been
      observed in~\citet{2008MNRAS.390.1470S}.
\end{itemize}

\begin{figure}
\centering
\includegraphics[width=8.5truecm]{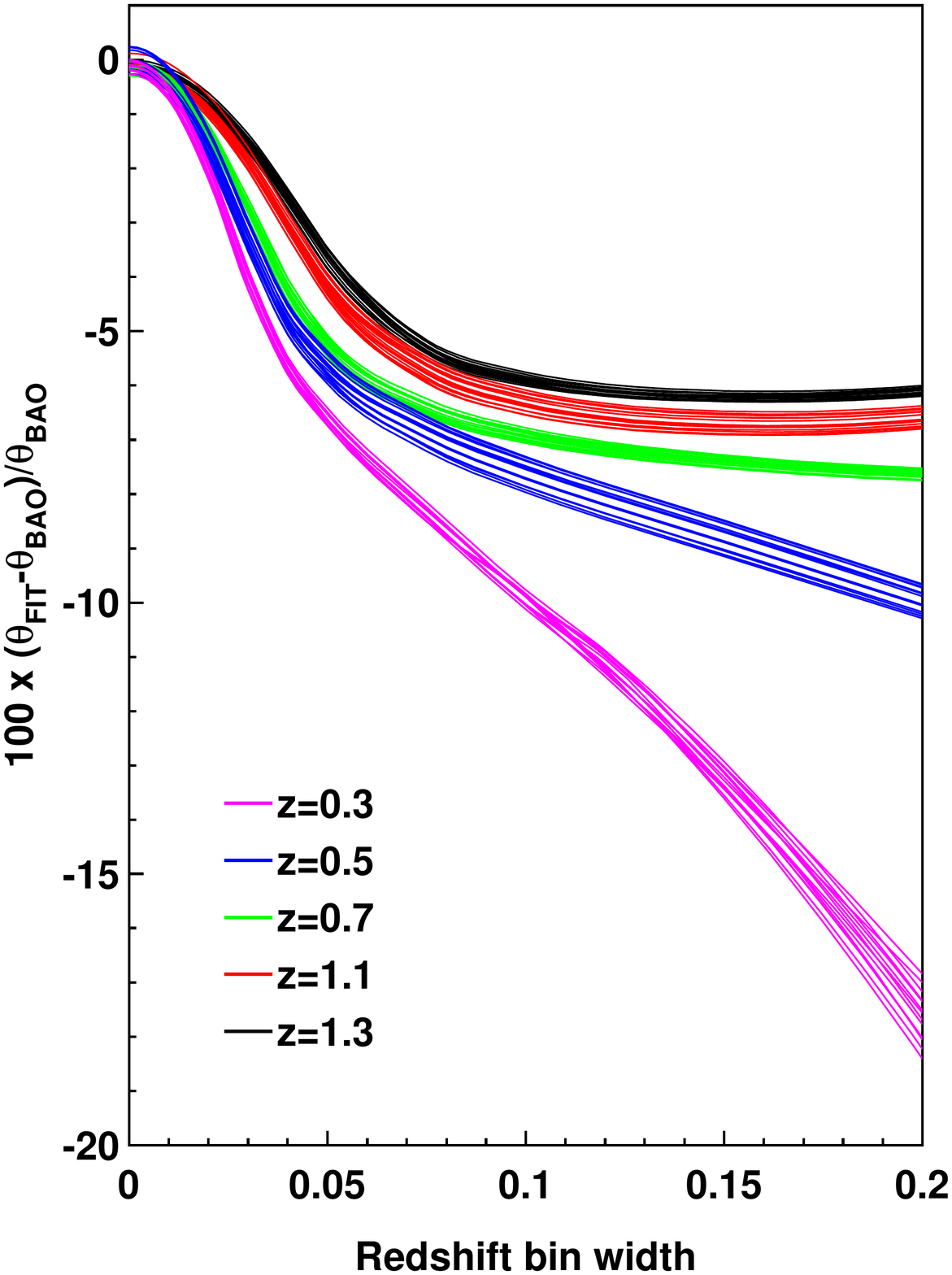}
\caption{Evolution of the shift of the measured $\theta_{FIT}$ with the
         width of the redshift bin for the 14 cosmological models considered
         in this work. The sound horizon scale $\theta_{BAO}$ for each
         cosmological model is taken as the reference. Two results are
         obtained. First, the correct sound horizon scale is recovered
         for any cosmological model within 0.75\% for infinitesimal redshift
         shells. Second, the shape of the shift is universal, and the spread
         is constant with the redshift bin width. Note that the dispersion 
         between models is much 
         smaller ($\leq 1\%$) than the shift due to increasing 
         binwidth. At low redshifts, $z<0.3$, the shift can be $\sim15$\% for 
         $\Delta z\geq0.1$, while at high redshifts, $z>0.5$, the shift 
         saturates at around 7\%.
         \label{fig:corr_indep_cosmo}}
\end{figure}

These two results are presented in Figure~\ref{fig:corr_indep_cosmo}, where 
we plot the evolution of
the shift, taking the angular scale corresponding to the sound
horizon scale as the reference, for 5 different redshifts, for 
several bin widths and for the 14 cosmological models. The spread
of the results is constant with the redshift bin width, and
comes from a possible residual dependence on the cosmology together
with the intrinsic limitations of the method (see 
section~\ref{subsec:results} on systematic errors). The average value of this 
shift is the correction applied to the fitted $\theta_{FIT}$. After this 
correction, the true value for the BAO scale is recovered for any
bin width and cosmology. Note that the absolute value of the projection effect
changes with the cosmological model, since both the BAO position and the
radial distance change. However, the relative effect is cosmology 
independent, as shown in  Figure~\ref{fig:corr_indep_cosmo}.

As has already been noted~\citep{2009PhRvD..79f3508S}, the projection 
effect is much more pronounced at low redshift. This must be taken into 
account to choose the optimum size of the redshift bins. The correction 
must be kept as small as possible, given the limitations imposed by the
photo-z precision, to introduce small systematic uncertainties. We are
using a Gaussian form for the photo-z uncertainty, because if the requirements
on photo-z measurement for a survey like DES are fulfilled, the effects 
of a possible non-Gaussianity are kept small. If this is not the
case, the evaluation of the true width of the bin must be refined. This
is another reason to maintain the correction small. 

\begin{figure}
\centering
\includegraphics[width=9.0truecm]{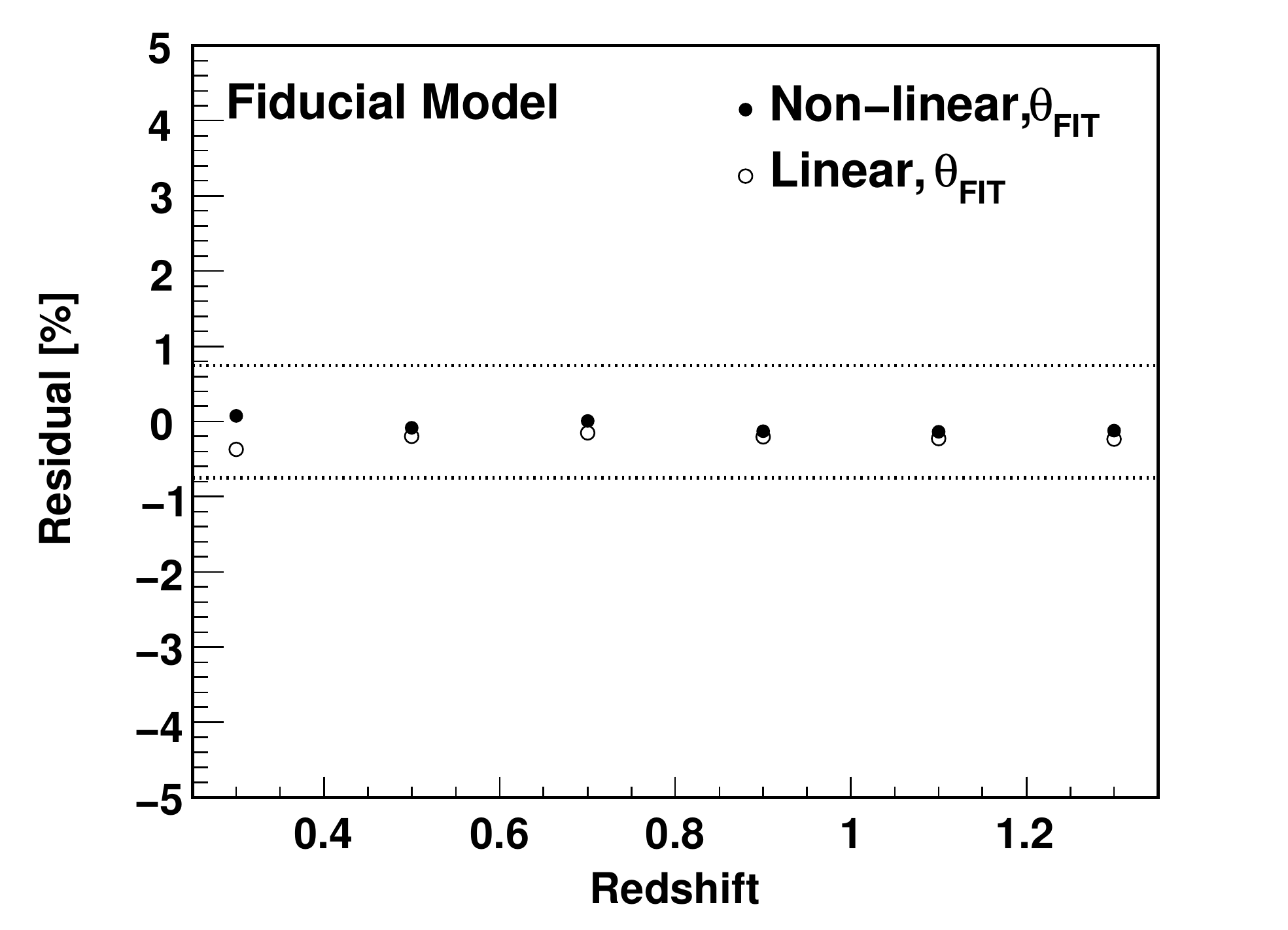}
\caption{Residual of the result of the parametric fit $\theta_{FIT}$ with respect 
         to the the theoretical value of the BAO scale as a function of redshift
         for the fiducial model. The open dots show the results obtained from the 
         fit to the linear correlation functions and the solid dots show the results
         on the non-linear correlation functions. In both cases the theoretical values
         are recovered, well inside the quoted systematic error of 0.75\%, shown by 
         the dotted lines. \label{fig:bao_lin_nonlin}}
\end{figure}

Note that the BAO scale is recovered both for linear and non-linear theory for
infinitesinal bin width. This can be seen in Figure~\ref{fig:bao_lin_nonlin}, where
the residual of the fit with respect to the theoretical $\theta_{BAO}$ as a
function of redshift is shown for the fiducial cosmological model. The recovered
values are well inside the 0.75\% precision that we quote as systematic 
error, represented by the dotted lines. This happens for all the cosmological 
models, showing that the method is robust and able to recover the theoretical 
BAO scale.

  \subsection{Redshift Space Distortions}
  \label{subsec:other_effects}

\newcommand{\xisp}{\xi(\sigma,\pi)}

There are other effects which are important in order to have under control
the parametric description of the angular correlation function. In 
particular, redshift space distortions must be taken into account in the
analysis of the BAO scale, since they result in an anisotropic correlation
function. We need to  replace $\xi(r)$ in Eq.~\ref{eq:xir2wtheta} 
by $\xisp$ with $\pi=\chi_2-\chi_1$, the radial separation, and 
$\sigma^2=2\chi_1\chi_2(1-\cos{\theta})$, the perpendicular
separation. Galaxy pairs separated by large radial distances
infall into each other  which means that they  
are measured  as contributing to 
bins of smaller radial separation (in redshift space) than in real (or
true) space. This results in larger correlations (more pairs)
at intermediate scales ($\pi<20$ Mpc/h) and lower correlation
at larger separations. The redshift space correlation becomes negative
for $\pi>40$ Mpc/h while the real space correlation remains positive
until $\pi>130$ Mpc/h (see Fig. A1 in \citealt{2009MNRAS.399.1663G}).
 Recall from Eq.~\ref{eq:xir2wtheta} how the
angular correlation is just a mean over the radial distance within the
corresponding redshift bin. This means that for narrow redshift bins (of 
width comparable to $\pi \simeq 100$ Mpc/h)  the resulting angular 
correlation can be quite different in real and redshift 
space~\citep{Fisher:1994cm,2007MNRAS.378..852P,Nock:2010gr}. As 
the total number of pairs is conserved, the 
integral over all separations is the same in real and redshift space, so 
that one expect to find the same angular correlation only for broad 
redshift bins, where the contribution of the boundary to the integral 
is negligible.
 
\begin{figure}
\centering
\includegraphics[width=0.50\textwidth]{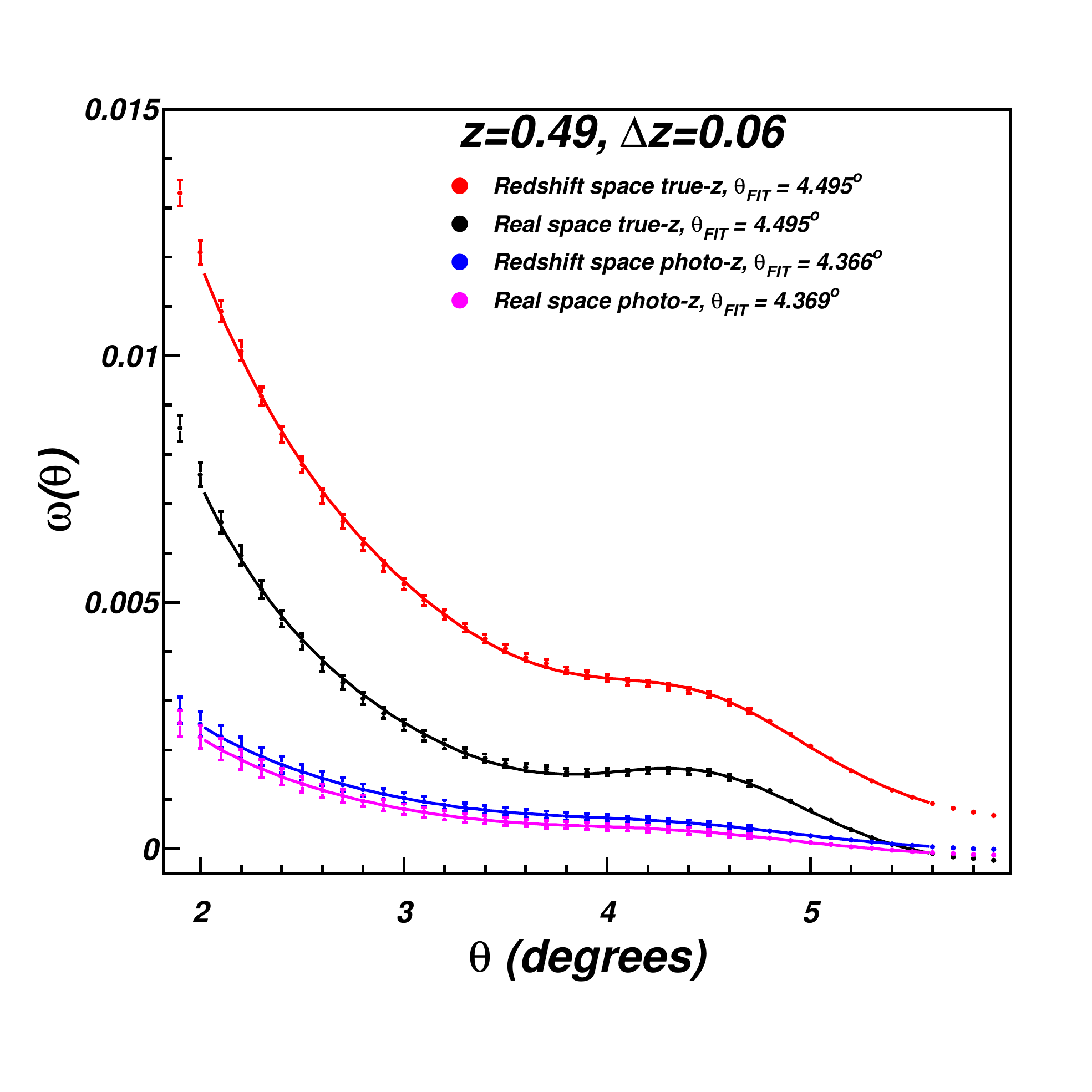}
\vskip-1.0truecm
\includegraphics[width=0.50\textwidth]{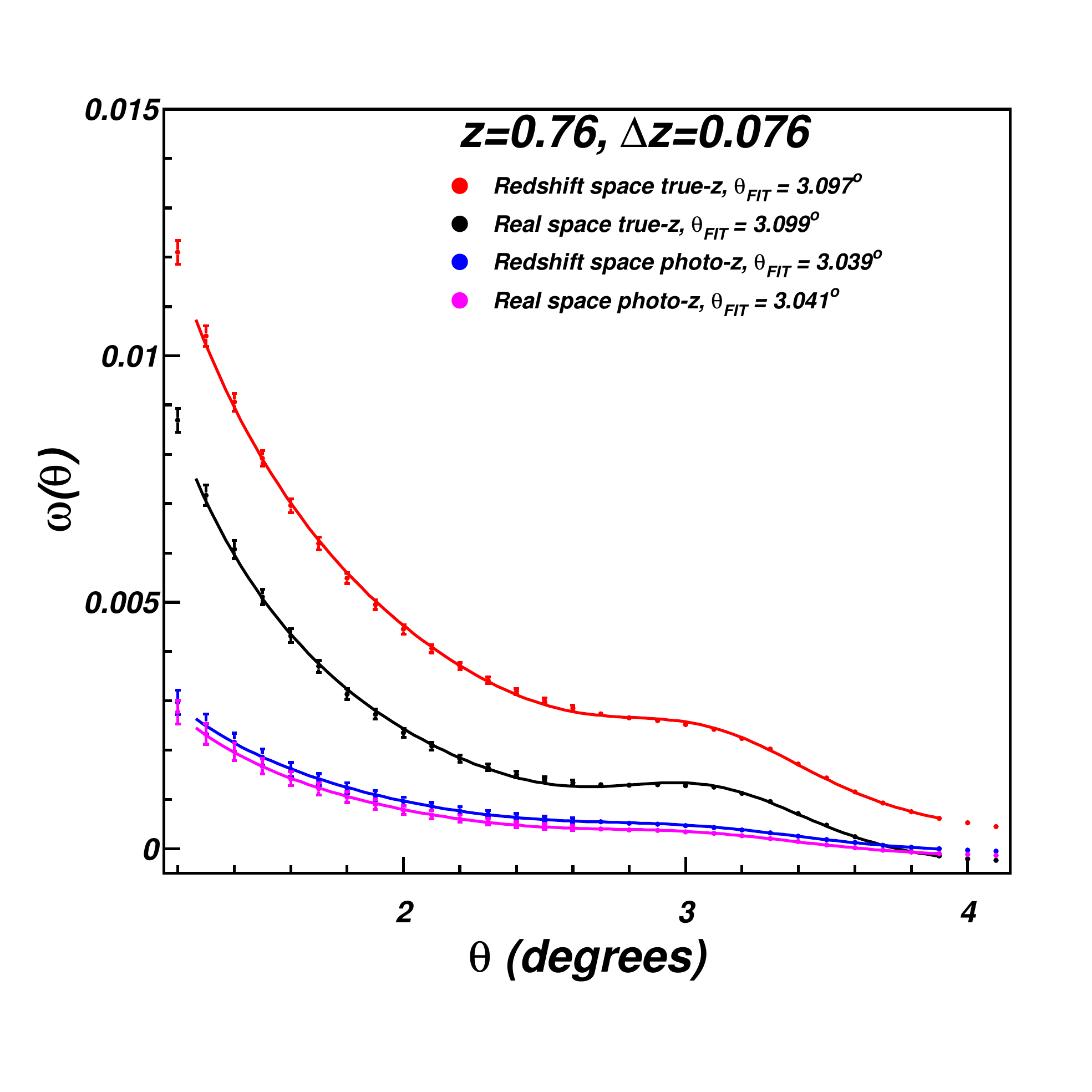}
\vspace*{-1.0truecm}
\caption{Effect of z-distortions and photo-z. Points (with nominal 2\% errors) 
show the linear theory prediction, while lines correspond to a fit to our 
parametrization in Eq.~\ref{eq:param}. Labels indicate the corresponding 
values of $\theta_{FIT}$ in each case.
\label{fig:z_dist}}
\end{figure}

The resulting predictions for redshift space agree very well with MICE
simulations (more details can be found in~\citealt{crocce10}).
Fig.~\ref{fig:z_dist} shows some examples of this redshift
space modeling for linear theory 
for two of the redshift bins used in the analysis presented above. 
Symbols (with 2\% nominal errors) correspond to the linear model
predictions when we include the redshift space distortions as well
as photo-z distortions. Note how both redshift space and photo-z
distortions change quite dramatically the amplitude of the 
resulting angular correlation.

The lines in Fig.~\ref{fig:z_dist} show the results of a fit
of our parametrization (in Eq.~\ref{eq:param}) to these new predictions. The 
corresponding fit values of $\theta_{FIT}$ are shown in the label. 
As expected, there is a small but systematic shift
of $\theta_{FIT}$ to smaller angles in the case of photo-z
distortions. This is because the effetive redshift bin is wider due to the 
photo-z errors. This effect is taken into account by the correction shown in 
Fig.~\ref{fig:corr_indep_cosmo} (once we use the equivalent width of 
the true redshift distribution).~\citet{crocce10} shows that the model 
we are using to predict the errors also works in redshift space.

As can be seen in Fig.~\ref{fig:z_dist}, the parametrization absorbs very well
the effects of redshift space distortions and the value recovered for
 $\theta_{FIT}$  agree well in real and redshift space despite the
 large difference in amplitude. This is not too surprising as we have
already shown above that this parametrization works for different 
cosmological models, different redshifts and also different redshift
widths.

\section{Application to the Dark Energy Survey}
\label{sec:applicationDES}

  \subsection{The Dark Energy Survey}
  \label{sec:des}
The Dark Energy Survey (DES)~\footnote{http://www.darkenergysurvey.org} is 
a next generation sky survey aimed directly at understanding the nature of 
the dark energy. It will use 
30\% of the available time on the Blanco telescope at CTIO, during 
5 years. The survey is designed to produce photometric redshifts in the 
range $0.2 < z < 1.4$, and will cover 5000 square degrees in the southern
hemisphere. The filters to be used are $g$, $r$, $i$, $z$, $Y$.

DES has been designed to exploit mainly 4 methods in order to
study the dark energy. These methods are the galaxy cluster counting and 
spatial distribution of galaxy clusters, the shifting of the 
galaxy spatial angular power spectra with redshift, weak gravitational 
lensing measurements on several redshift shells, and distances
to $\sim 2000$ supernovae. Using these techniques, a
5\%-15\% precision measurement in the equation of state parameter $w$ 
from each of our methods, and a 30\% measurement in the variation of 
$w$ with time are expected. Combined, they provide both stronger 
constraints and a check on systematic errors.

  \subsection{N-body Simulation Features}
  \label{sec:simulation}
We have developed and tested our method to recover the angular scale
of the sound horizon using a large N-body simulation capable of
reproducing the geometry (e.g. area, density and depth) and specifications 
of DES~\footnote{This catalog is publicly available at http://www.ice.cat/mice}.

The simulated data was kindly provided by the MICE project 
team, and consisted of a distribution of dark 
matter particles (galaxies, from now on) with the cosmological parameters
fixed to the fiducial model of Table~\ref{tab:cosmomodels}. The radial
profile matches the DES expectation

\begin{equation}
dN/dz \propto (z/0.5)^2 \exp{-(z/0.5)^{1.5}},
\end{equation}

\noindent
the simulation covers $1/8$ of sky ($\sim 5000$ square degrees) in the 
redshift range $0.2 < z < 1.4$, containing 50 million galaxies. All 
these numbers roughly match the DES expectations for the main 
galaxies. These data 
was obtained from the comoving output at $z=0$ of one of the largest 
N-body simulations completed to date, with comoving size 
$L_{box}=3072\,{\it h}^{-1}\,{\rm Mpc}$ and more than $8\times 10^9$ 
particles ($m_p=2.3 \times 10^{11}\,{\it h}^{-1}\,M_{\odot}$). 
More details about this run can be found in \citet{2008MNRAS.391..435F}
and \citet{2009arXiv0907.0019C}.

\begin{figure}
\centering
\includegraphics[width=0.50\textwidth]{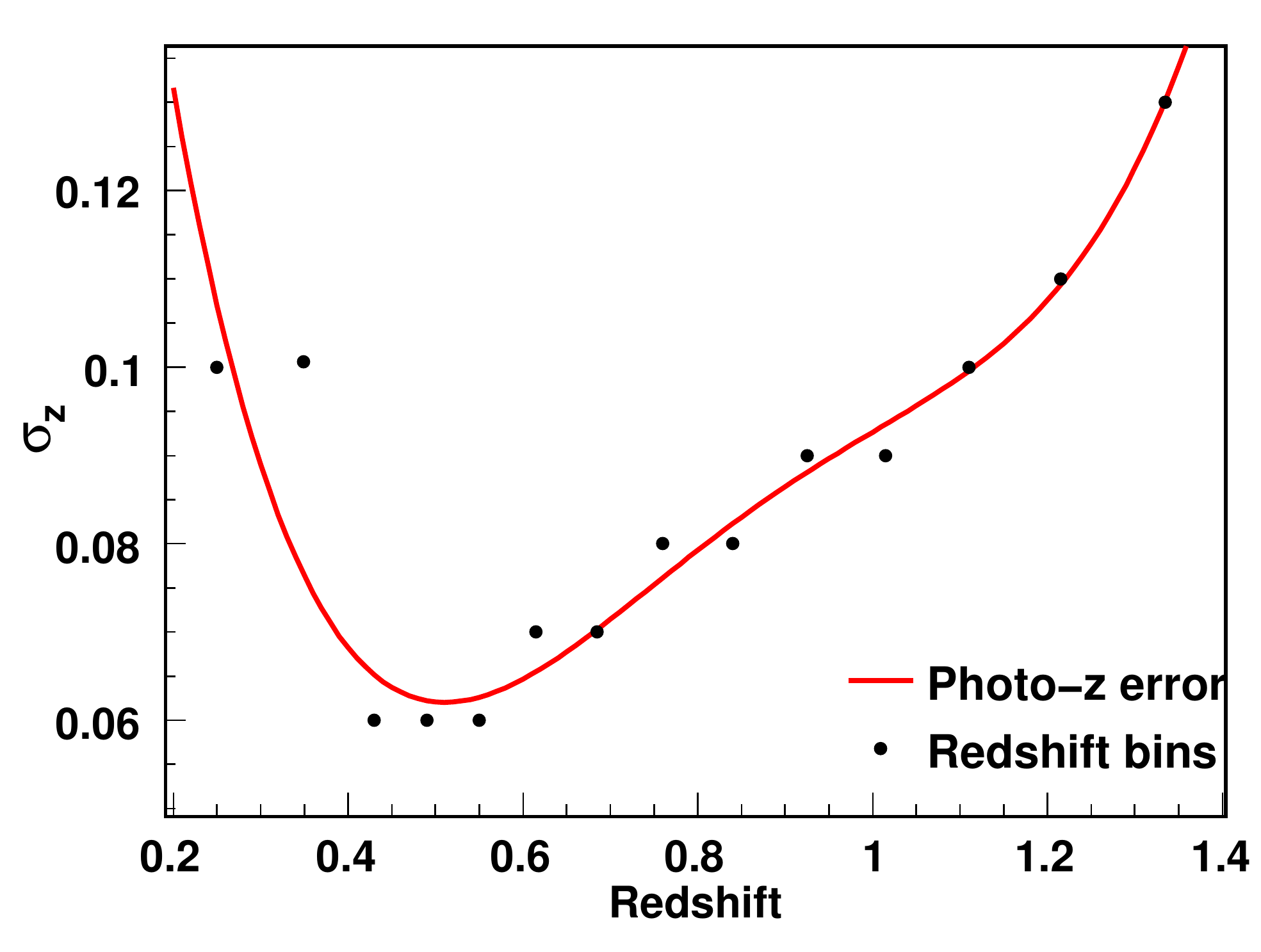}
\caption{Photo-z precision used in this analysis (continuous line), as
         compare to used bins position and width (dots).\label{fig:DESphotoz}}
\end{figure}

The precision in the determination of the redshift for galaxies 
which has been used in this analysis is shown in 
Figure~\ref{fig:DESphotoz}. This is the typical precision that is
obtained using the current photo-z codes for 
DES~\citep{2008MNRAS.386.1219B}. This photo-z has been introduced into
the simulation, to produce a catalog of galaxies similar to that expected
in DES, where the proposed procedure to recover $\theta_{BAO}$ has been 
applied. The photo-z has been introduced by scattering each individual redshift 
according to the Gaussian representing the photo-z error of 
Figure~\ref{fig:DESphotoz}. The redshift bin widths have been 
chosen as a compromise between statistics, expected precision in the photo-z 
determination and shift of the BAO peak due to projection effect. Then, the 
redshift bins used in this analysis follow the expected precision
in the photo-z measurement, being wider 
for low and high redshifts and narrower for mid redshift. The analysis is 
performed in the full redshift range, from $z=0.2$ to $z=1.4$. With 
this prescription we have 14 bins, represented
also as dots in Figure~\ref{fig:DESphotoz}. The position of the dot in
the horizontal axis represents the bin center, and the position in the
vertical axis represents the bin width, which has been chosen to be
close to the expected $1-\sigma$ error in photo-z measurement. Much wider bins
make the sensitivity too small due to projection effect (the amplitude
of the correlation function becomes very small and the shift of the
peak becames very large), while much narrower bins make the correlations
between redshift bins too large, complicating the analysis and also 
diminishing the sensitivity.

  \subsection{Results}
  \label{subsec:results}
\begin{figure*}
\centering
\leavevmode
\begin{tabular}{ccc}
\includegraphics[width=0.33\textwidth]{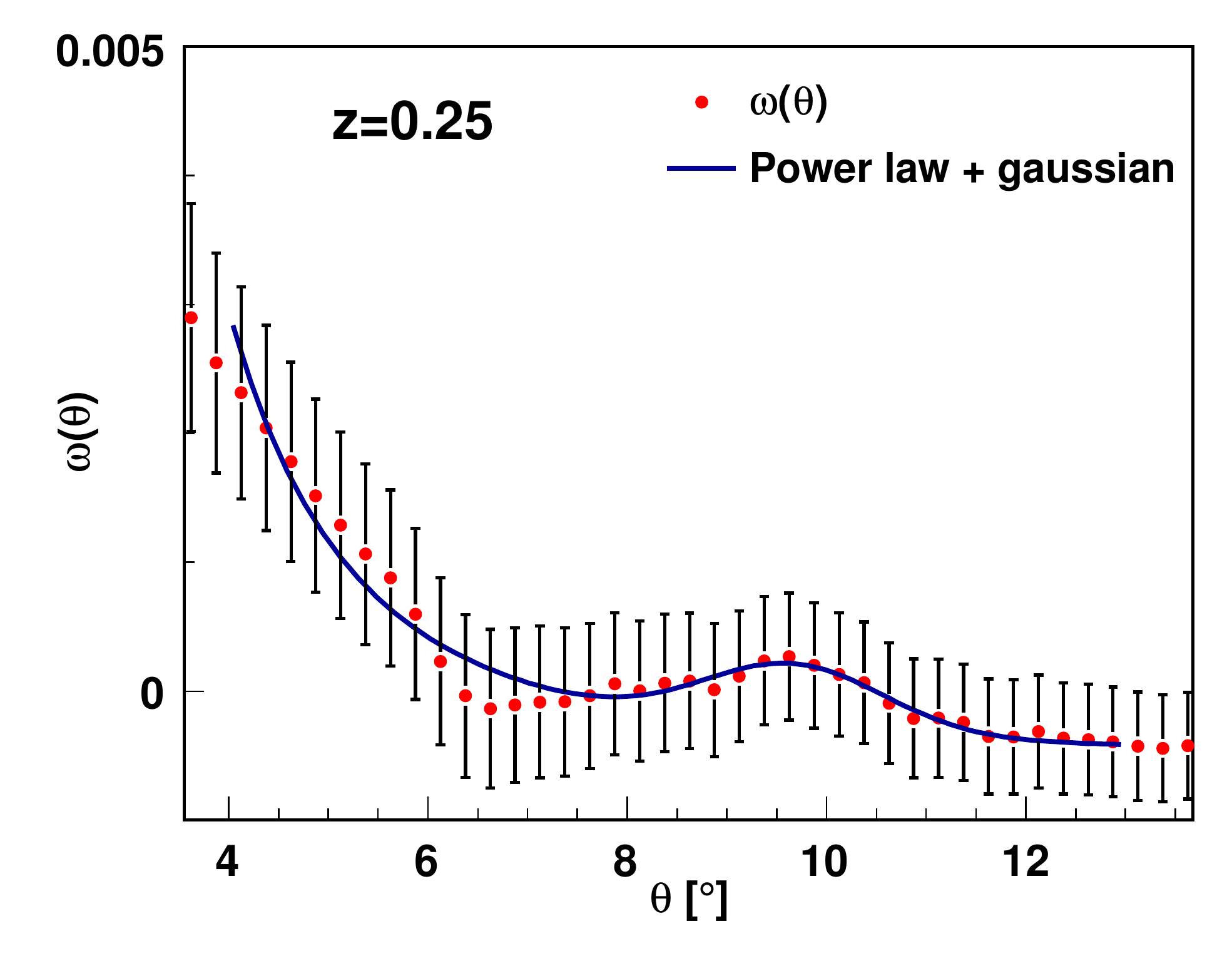} &
\includegraphics[width=0.33\textwidth]{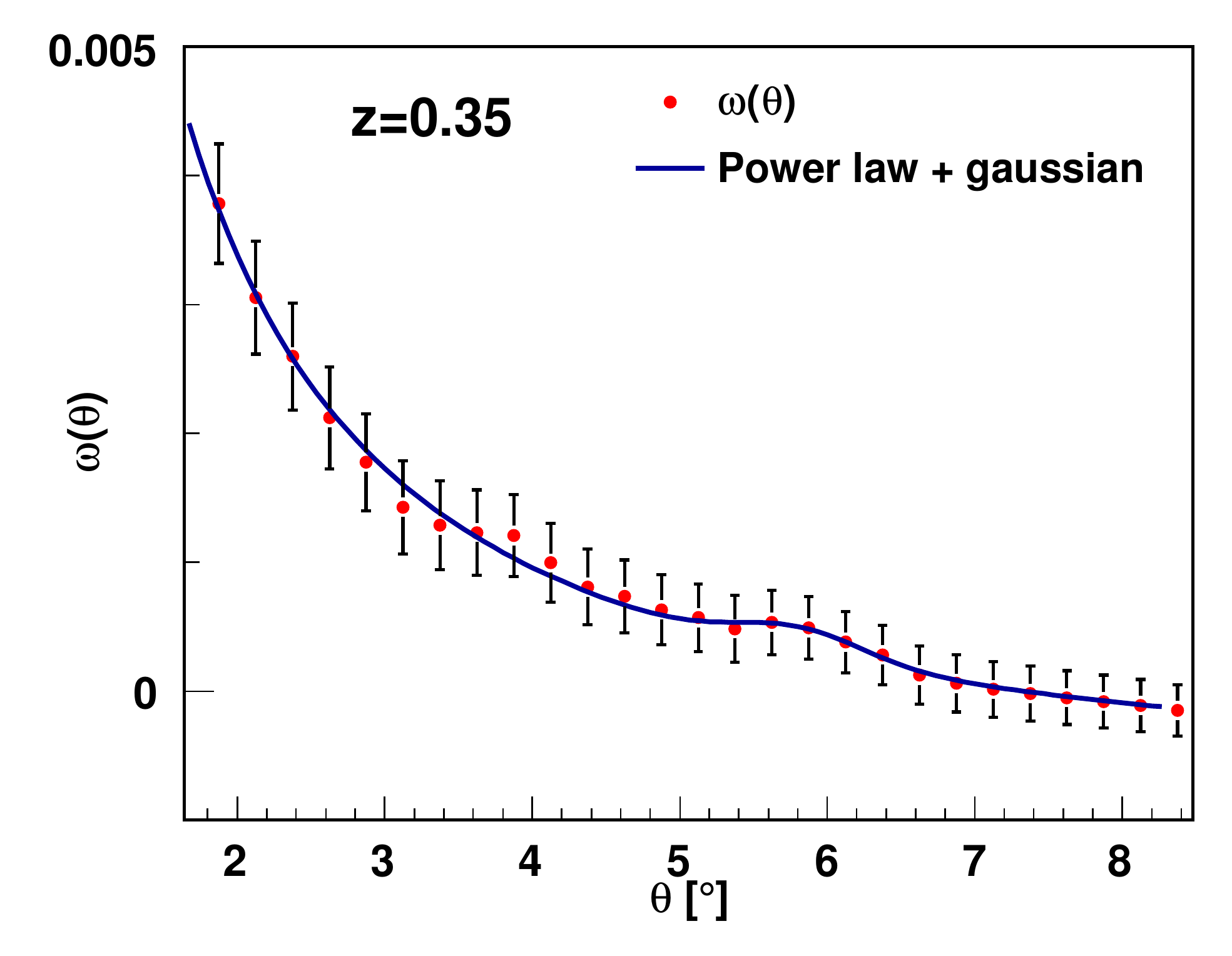} &
\includegraphics[width=0.33\textwidth]{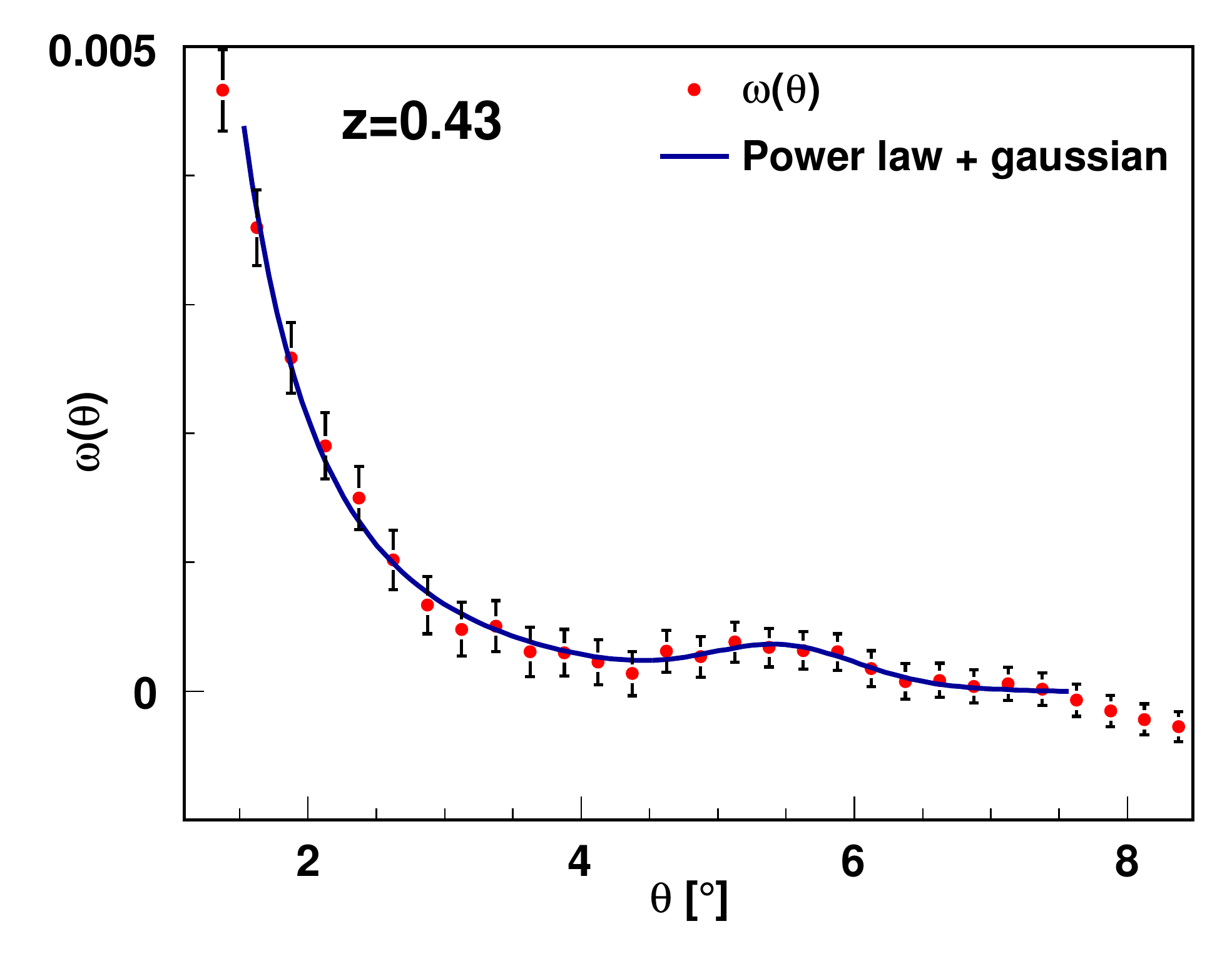} \\
\includegraphics[width=0.33\textwidth]{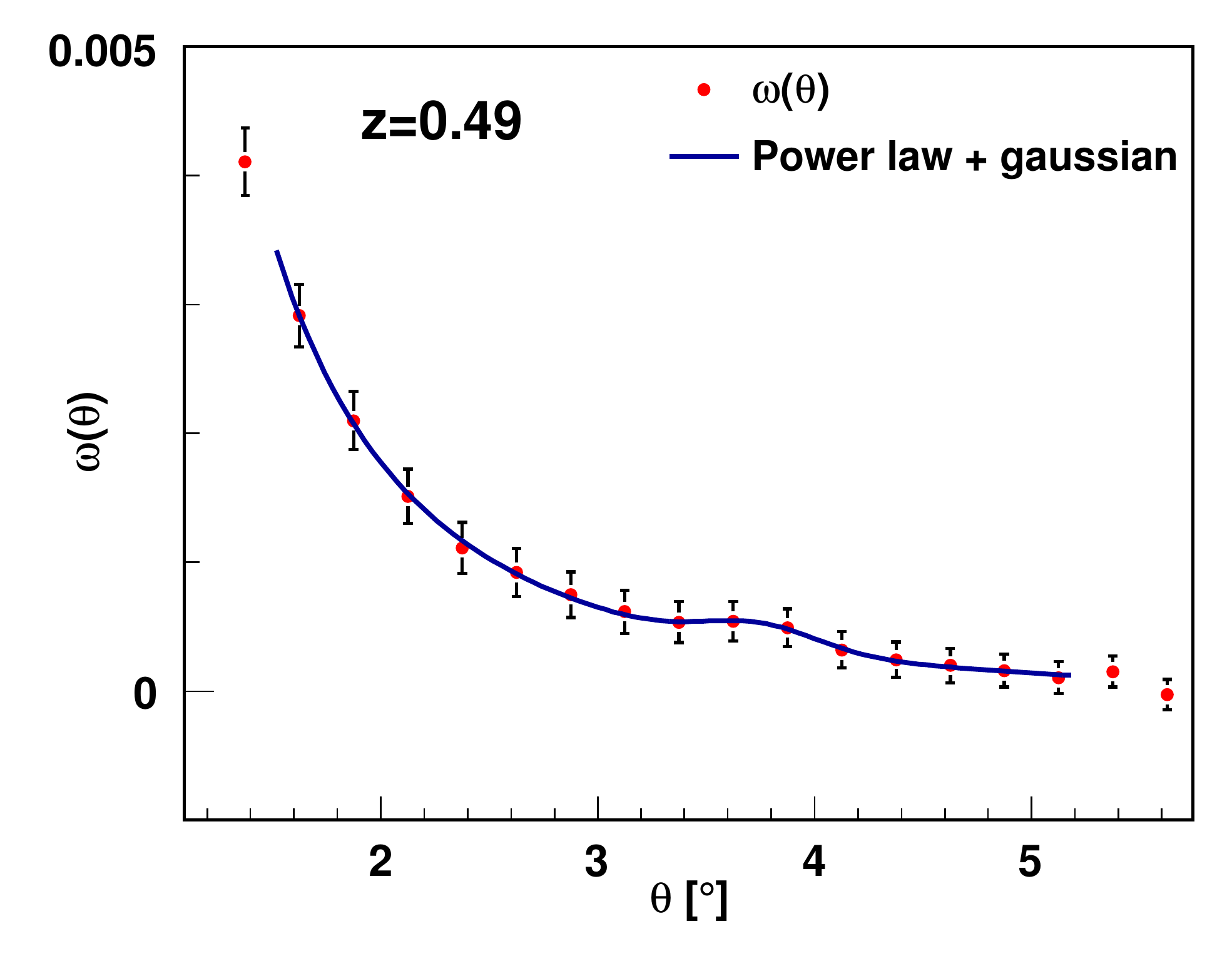} &
\includegraphics[width=0.33\textwidth]{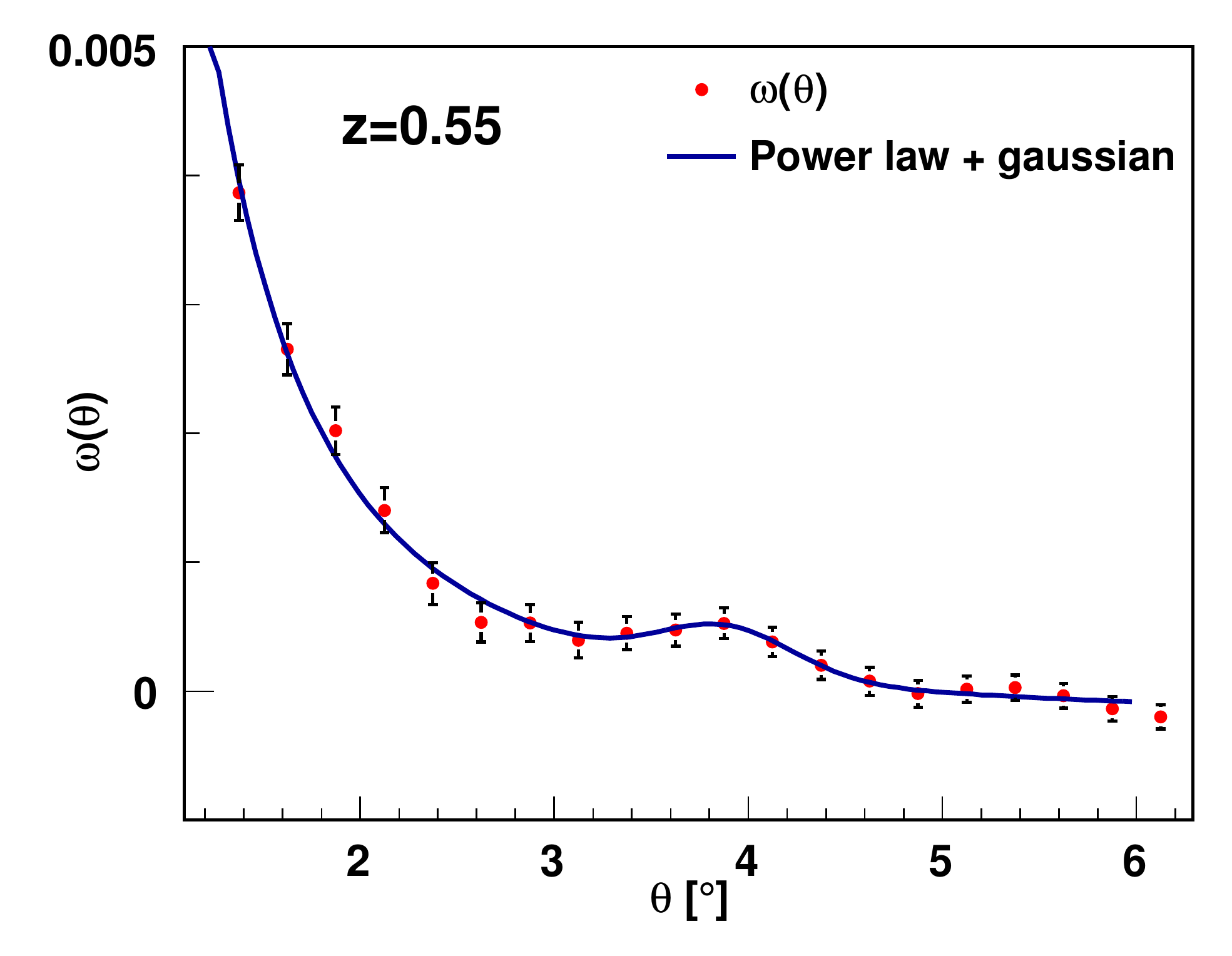} &
\includegraphics[width=0.33\textwidth]{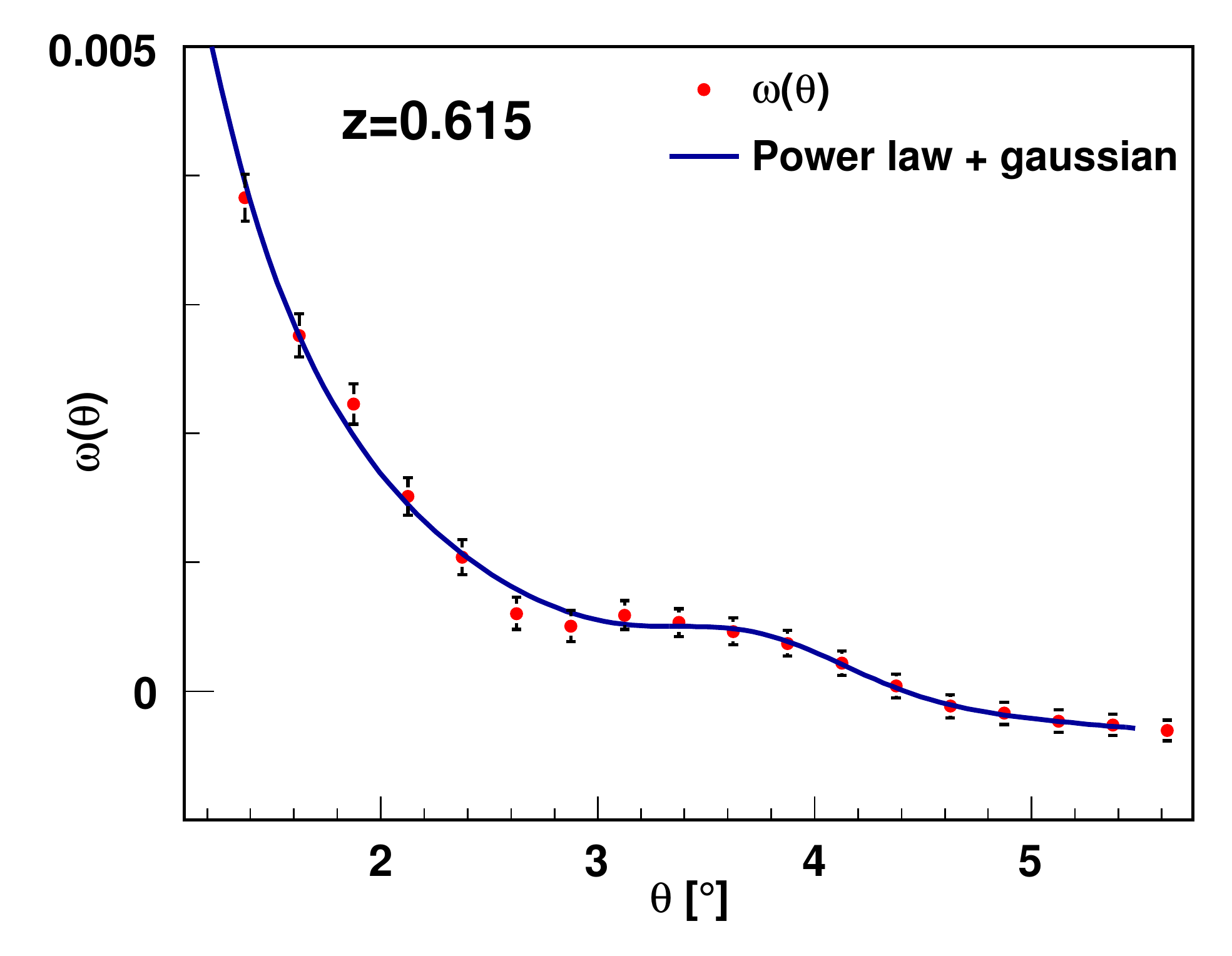} \\
\includegraphics[width=0.33\textwidth]{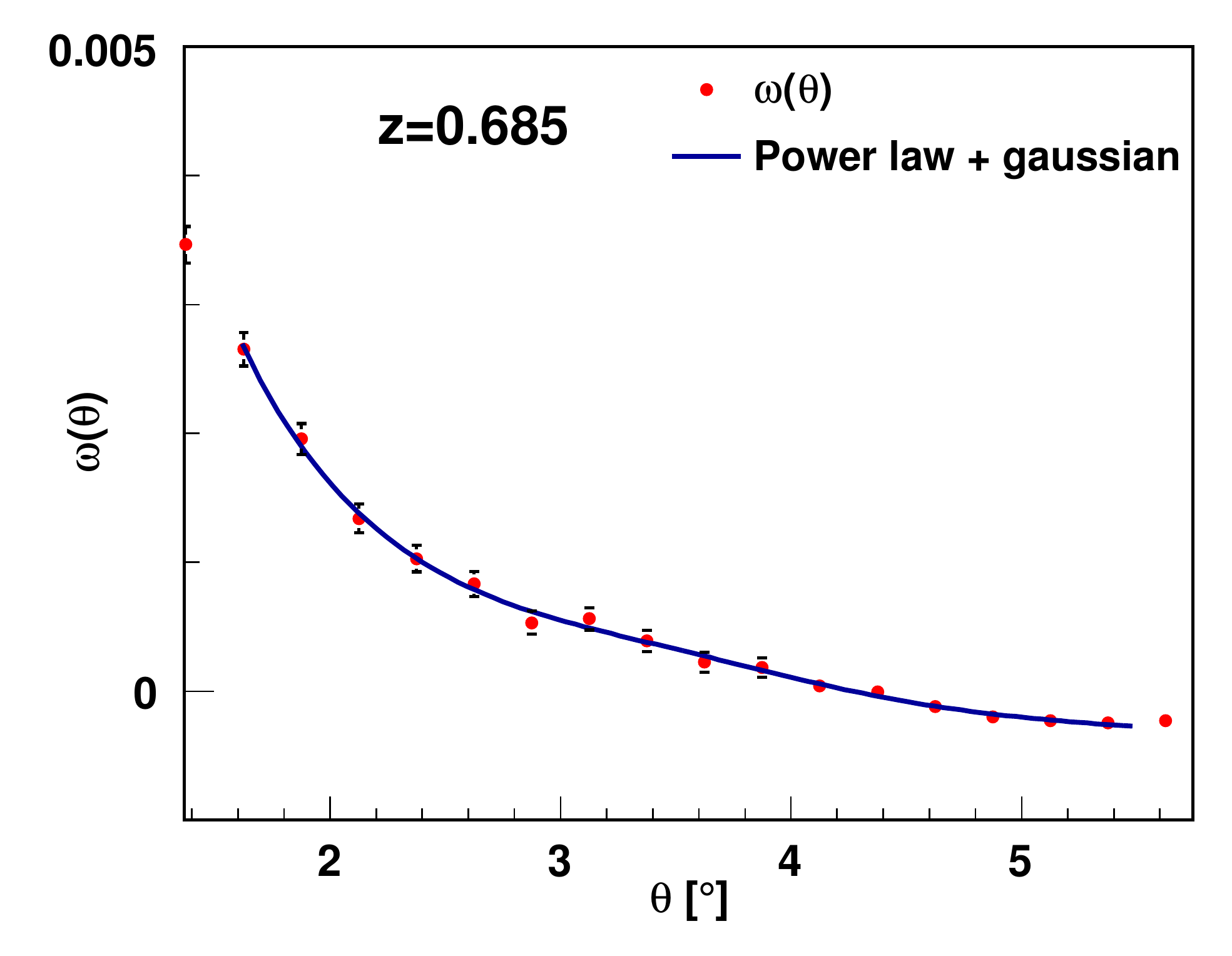} &
\includegraphics[width=0.33\textwidth]{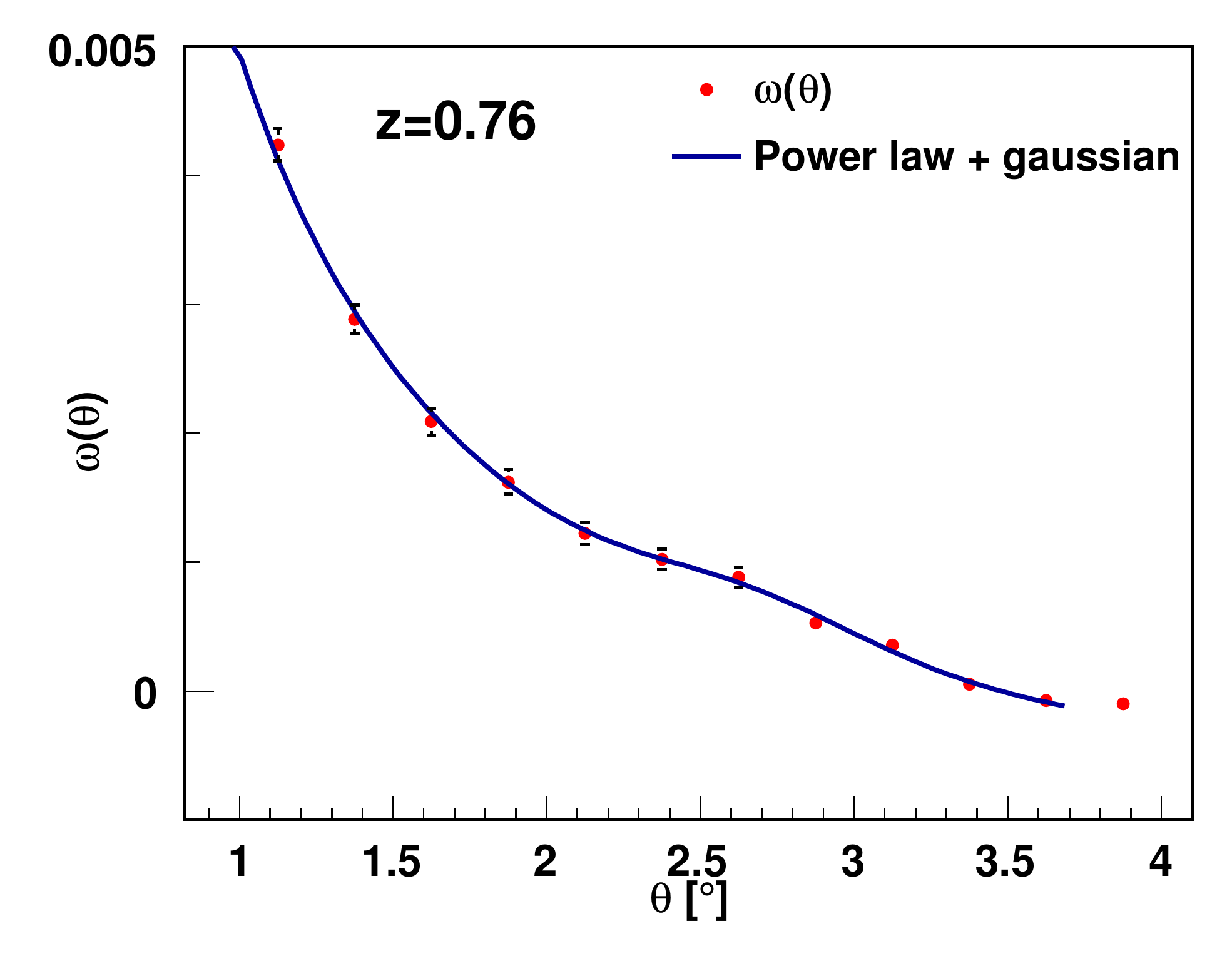} &
\includegraphics[width=0.33\textwidth]{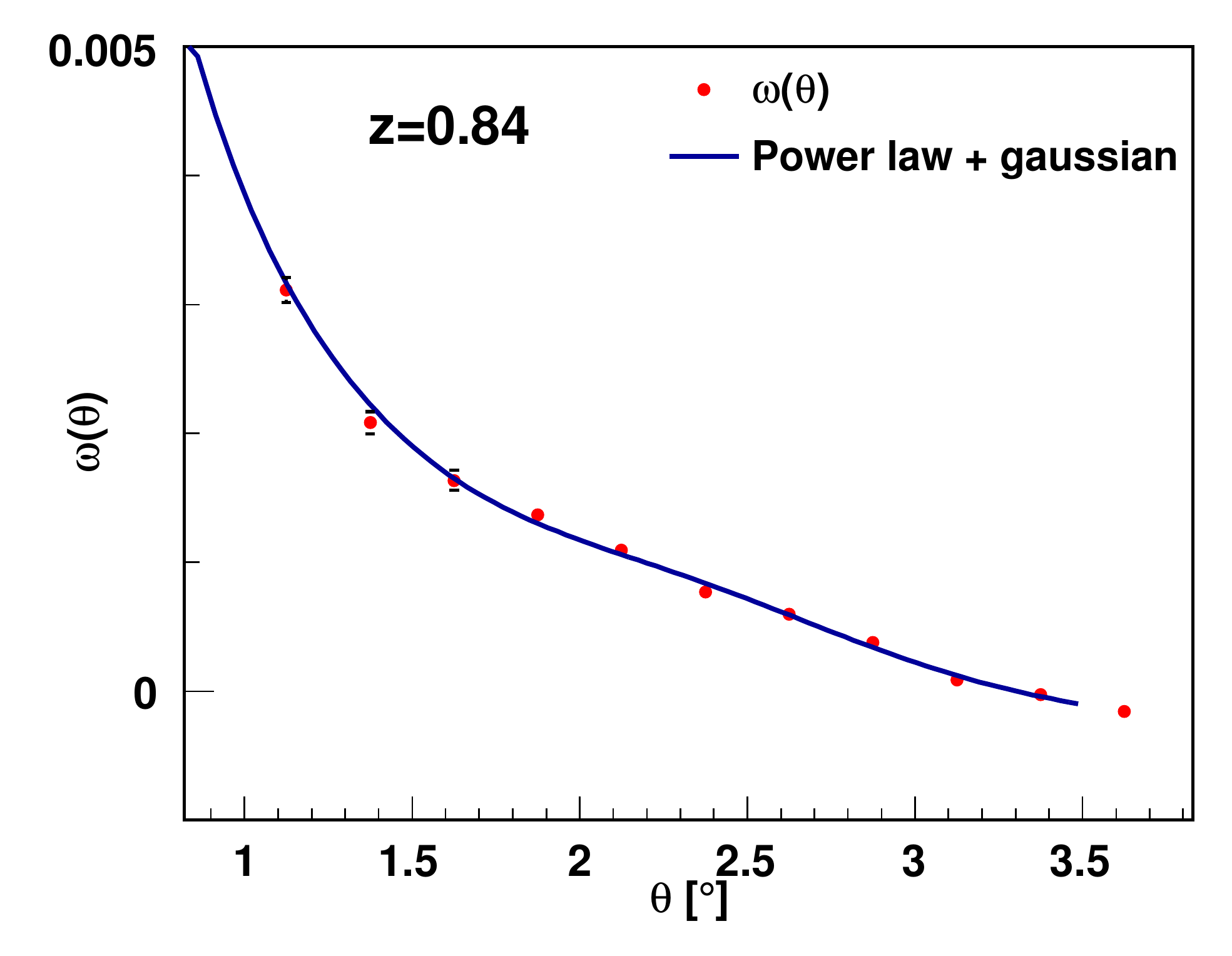} \\
\includegraphics[width=0.33\textwidth]{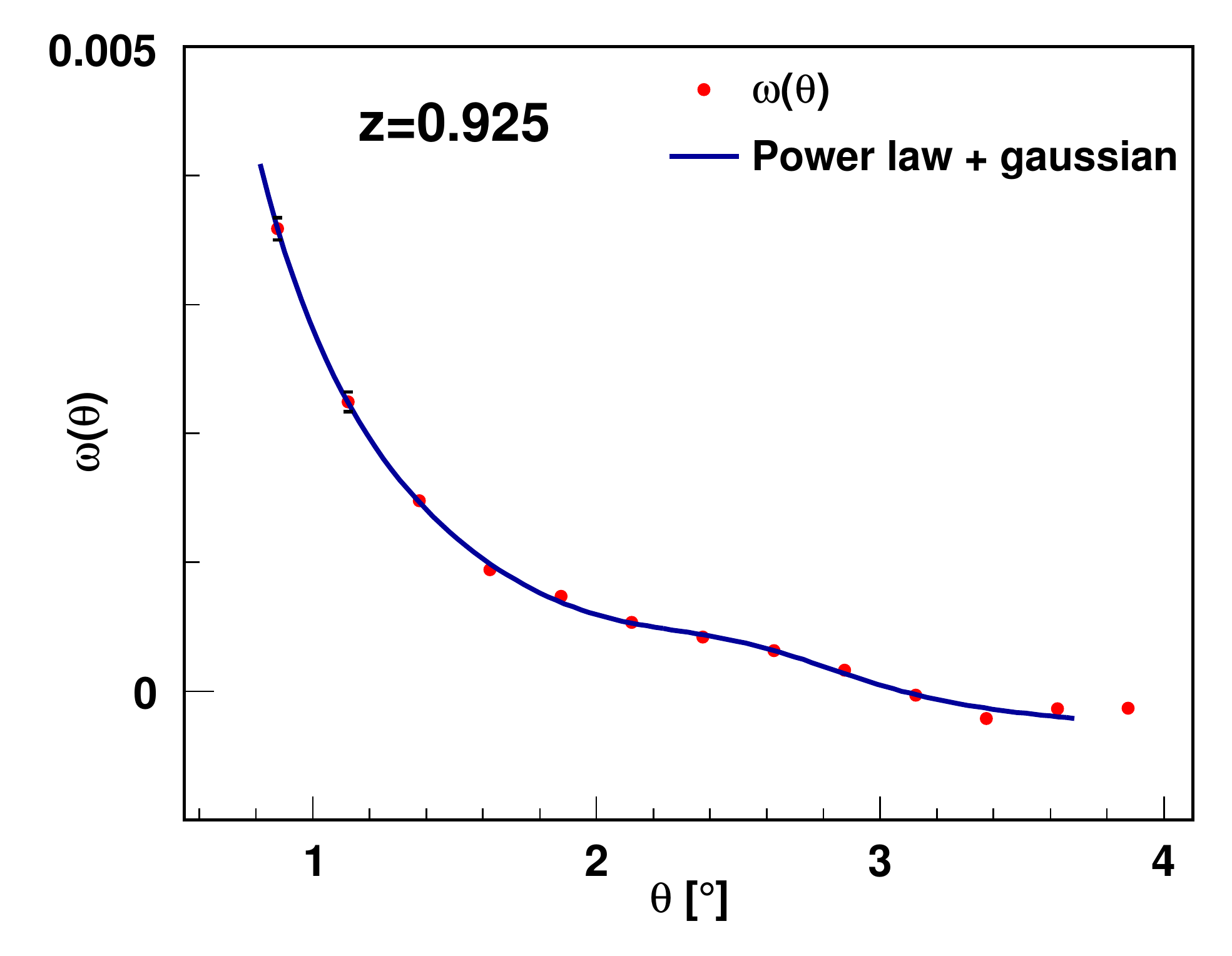} &
\includegraphics[width=0.33\textwidth]{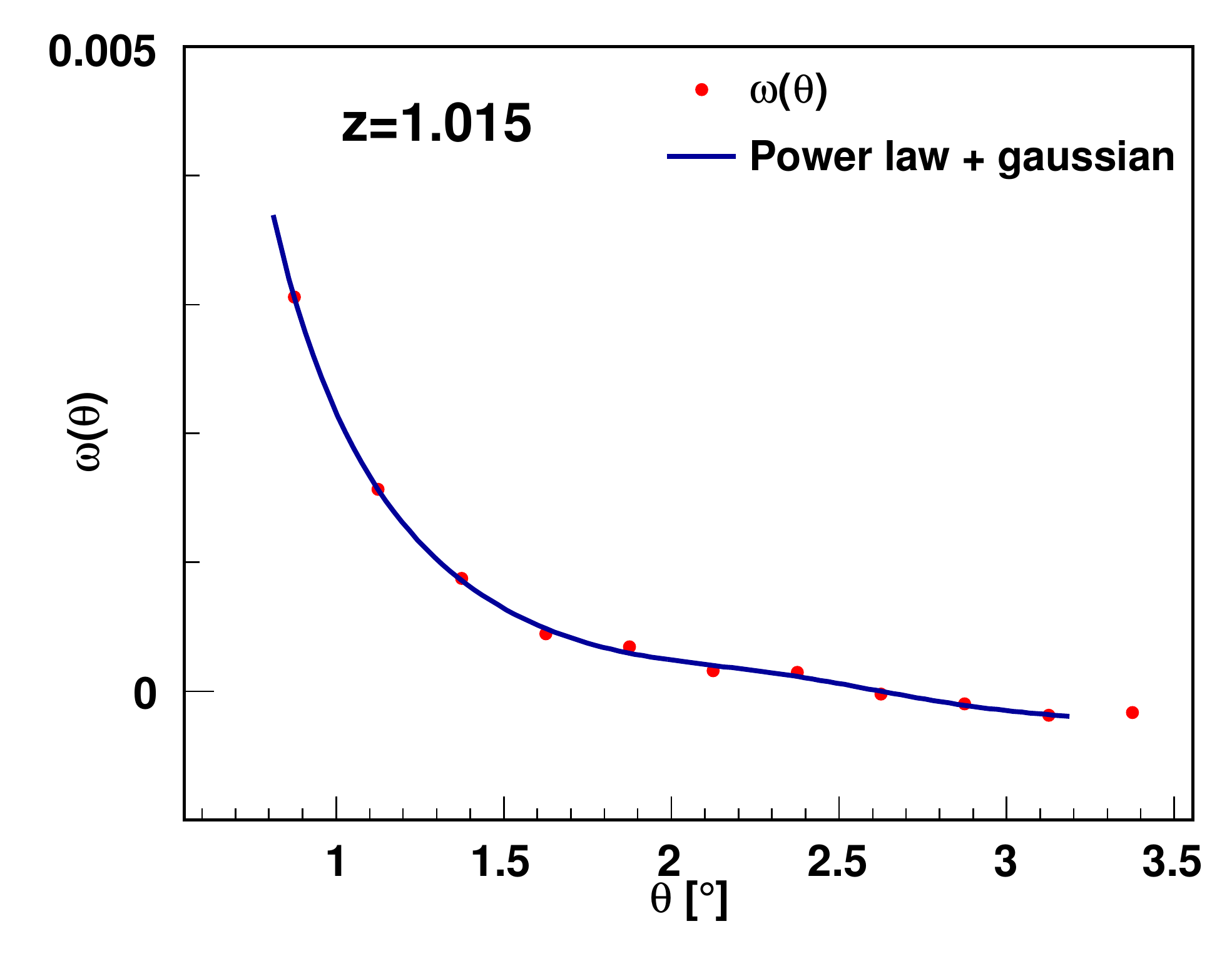} &
\includegraphics[width=0.33\textwidth]{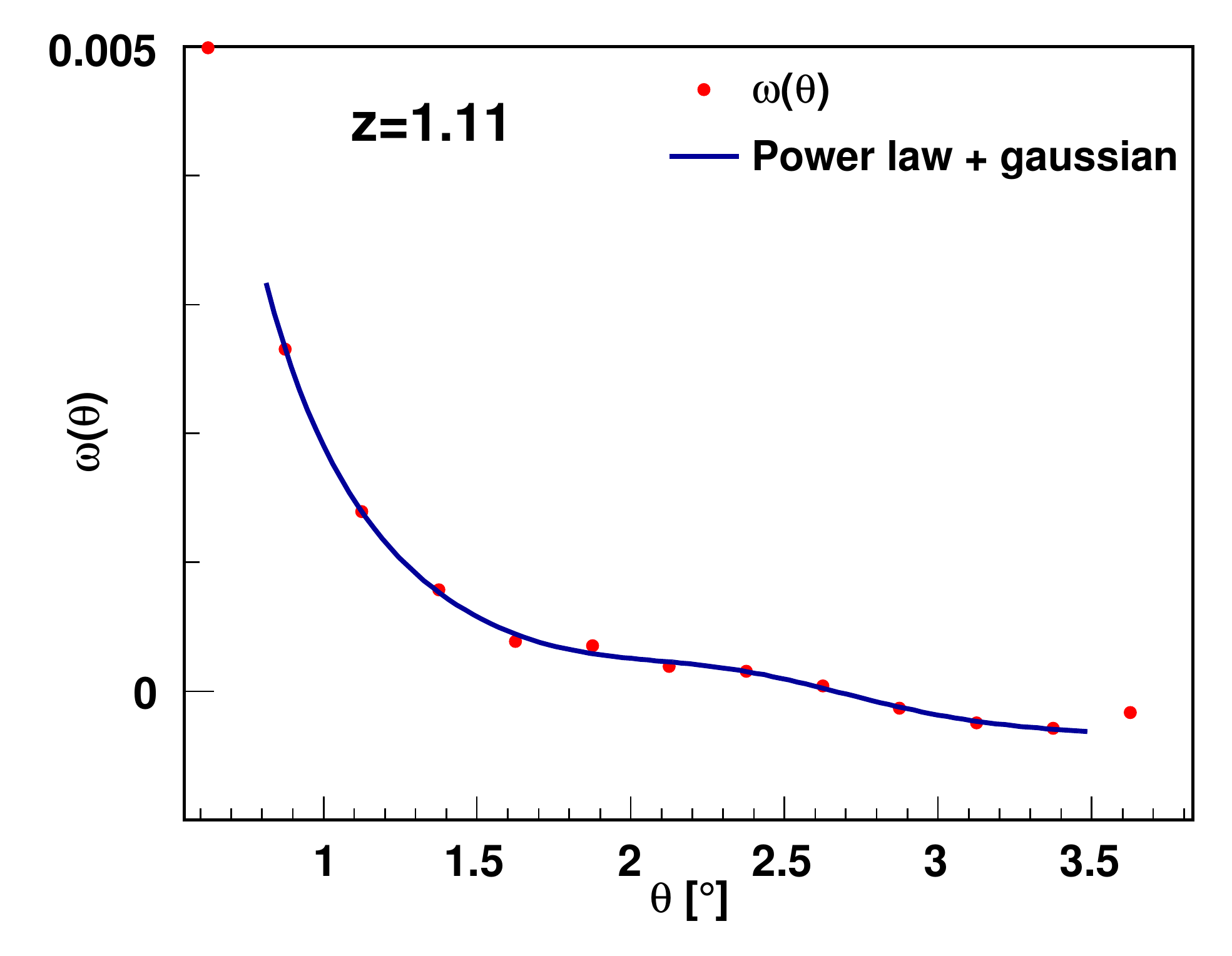} \\
\includegraphics[width=0.33\textwidth]{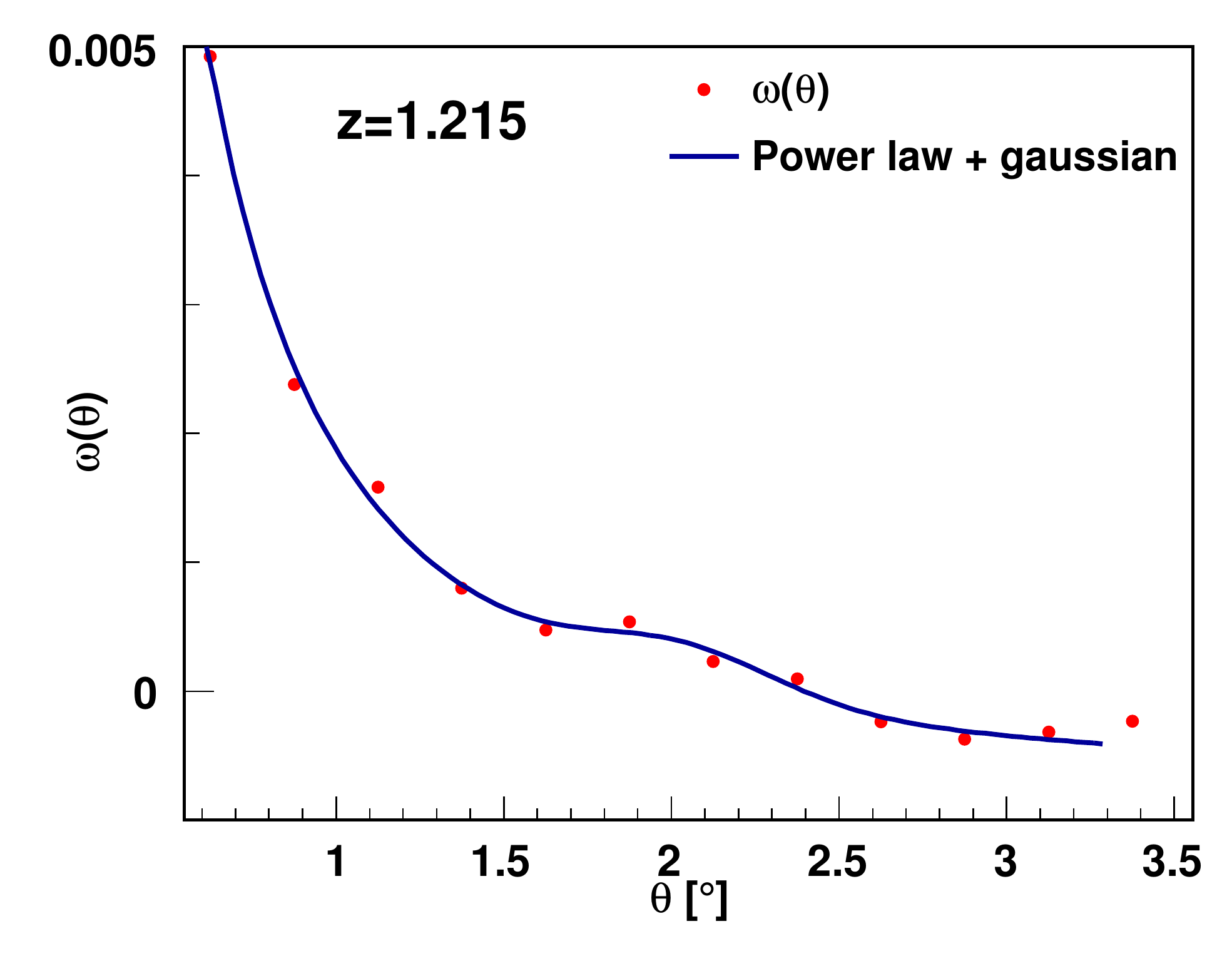} &
\includegraphics[width=0.33\textwidth]{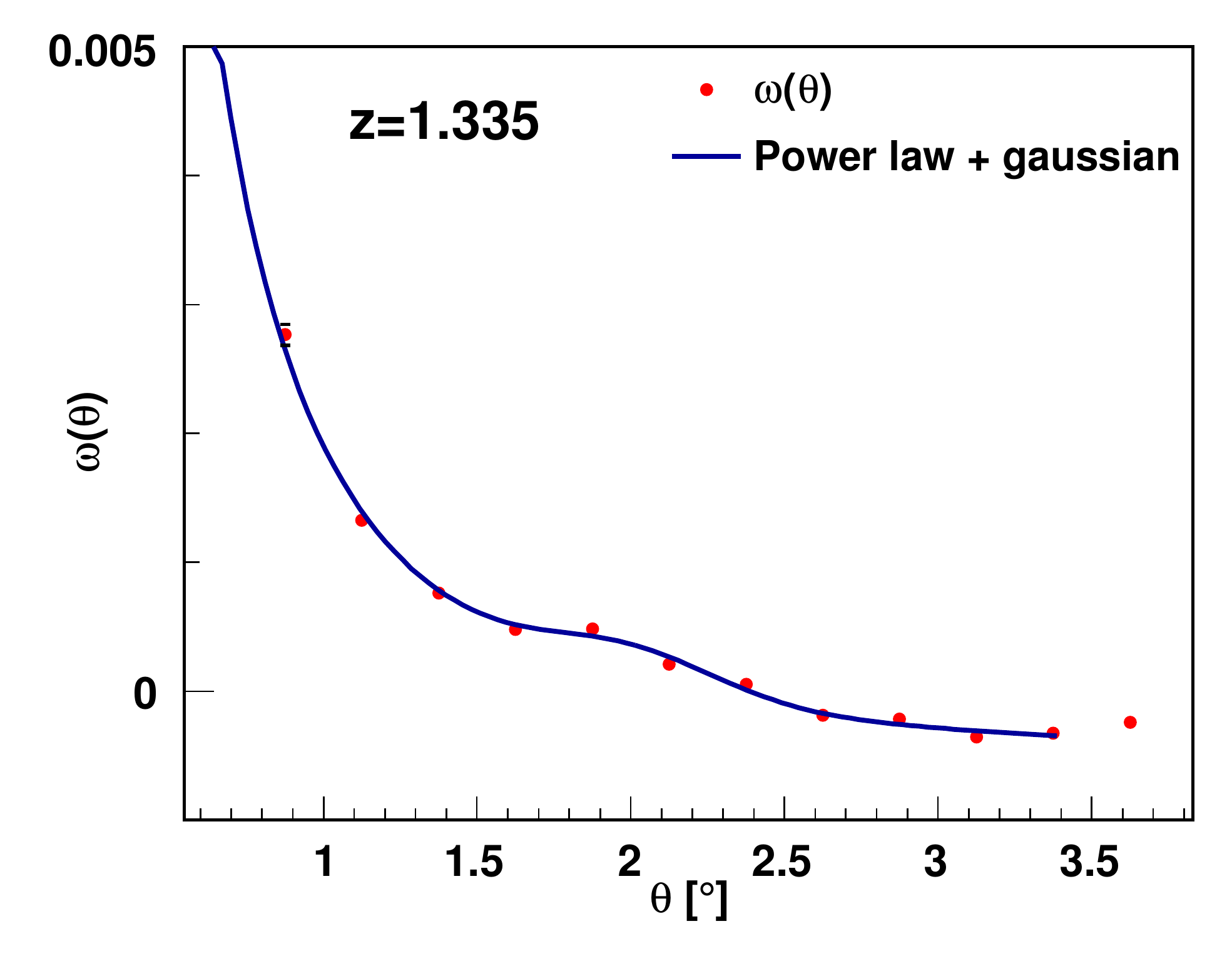} &
~ \\
\end{tabular}
\caption{Correlation functions for MICE simulation, including photo-z 
         effects. The errors are computed using 
         Eq~\ref{eq:cov}.\label{fig:cfmice1}}
\end{figure*}

In order to numerically estimate the angular correlation function in a
given redshift bin, we have used the Landy-Szalay
estimator \citep{1993ApJ...412...64L}. The random catalogs contain as many 
galaxies as the simulation, since the shot noise error is negligible, given 
the size of the sample.

The resulting correlation functions together with their parametrizations
are presented in Fig.~\ref{fig:cfmice1}. The statistical error 
in $\theta_{FIT}$ is given as the error in the
fit coming from the error in the correlation function, as explained in
section~\ref{sec:theoryAngCor}, and taking into account the photo-z effect, as
described in Eq.~\ref{eq:cov} below.

\begin{figure}
\centering
\includegraphics[width=0.50\textwidth]{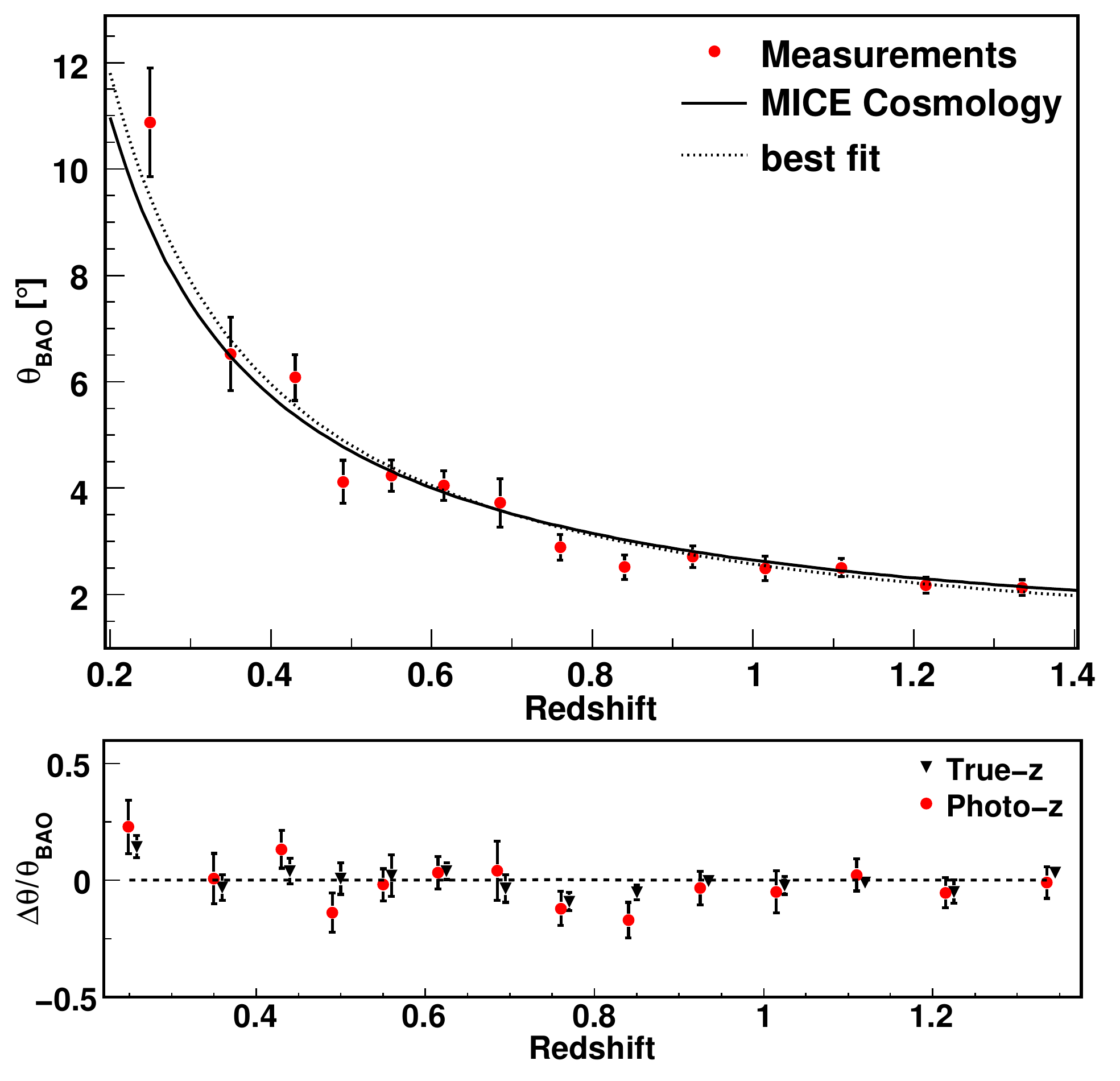}
\caption{Evolution of the measured $\theta_{BAO}$ with the 
         redshift (top). Comparison of the results obtained using
         photo-z and the true-z (bottom). The true-z results are closer
         to the MICE cosmology, as expected.\label{fig:thbao_vs_z}}
\end{figure}

The recovered values of $\theta_{BAO}$ after applying the correction
of Eq.~\ref{eq:corrtheta} can be seen 
in Fig.~\ref{fig:thbao_vs_z} (top) as a function of the redshift. The
correction is applied after evaluating the true width of the redshift
bin, using $\Delta z_{true} = \sqrt{2 \pi} \Delta z_{photo}$, which 
corresponds to the width of a top-hat bin in true-z which gives the 
same amplitude in $\omega(\theta)$ than the chosen photo-z bin. The points
include also the systematic errors, described in section~\ref{subsec:sys}. The 
MICE cosmology is also shown as the solid line, and the best fit
cosmology as the dashed line. To compare, the same analysis has
been applied to the true redshifts catalog ({\it i.e.}, without 
photo-z). The corresponding 
result is presented in the bottom panel of 
Figure~\ref{fig:thbao_vs_z}, together with the photo-z. The results
obtained using the true redshift are closer to the MICE cosmology, as
expected. For true redshifts, the bin width is not corrected, and the
systematic errors do not include any photo-z contribution.

\begin{figure}
\centering
\includegraphics[width=0.50\textwidth]{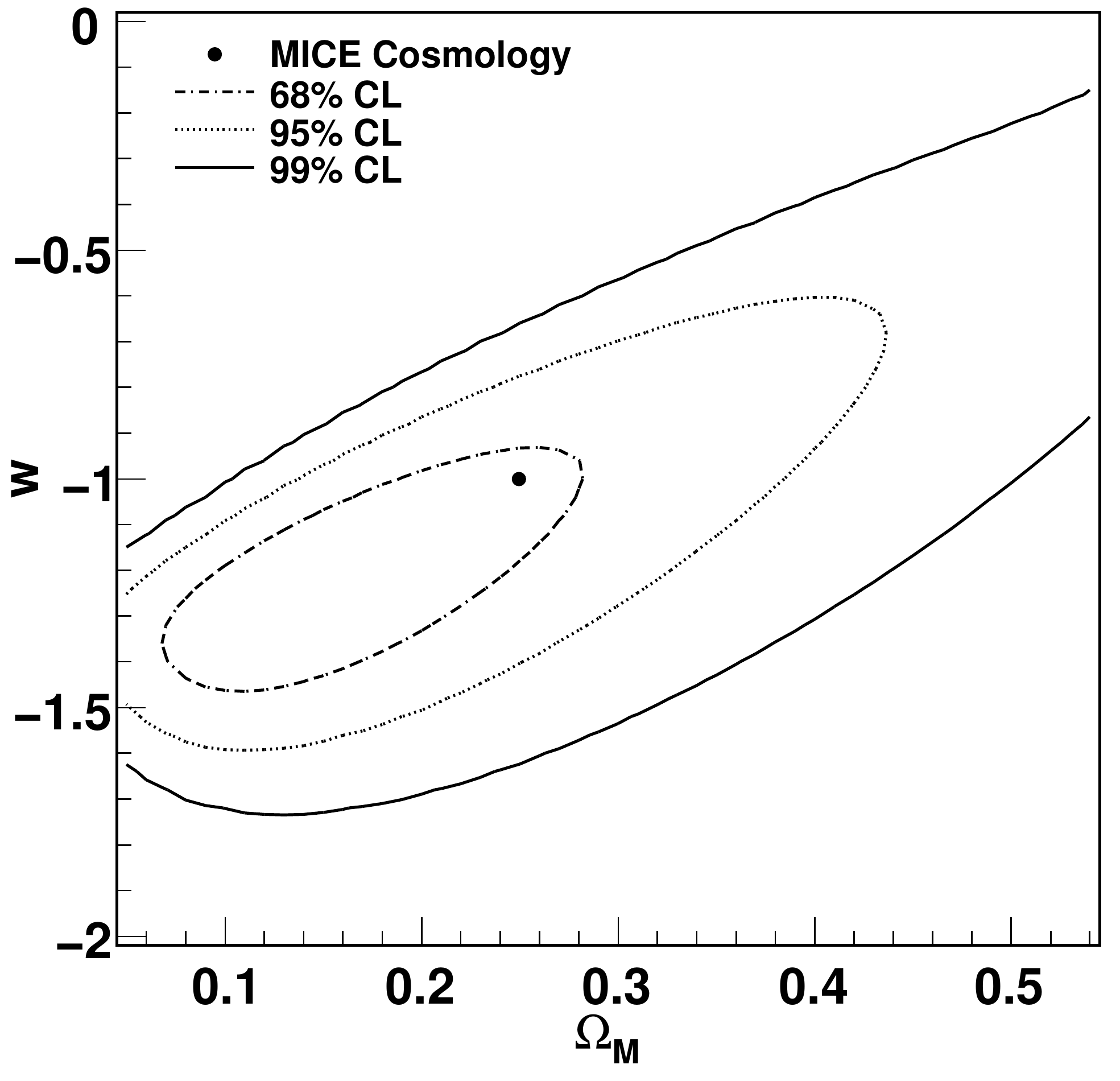}
\caption{Allowed region for the cosmological parameters at 68 
(dashed-dotted line), 95 (dashed line) and 99 (solid line) \% C. L. The 
true cosmology, given by the dot, is recovered within 
1-$\sigma$ .\label{fig:omvsw}}
\end{figure}

The true cosmology 
is recovered after applying the analysis method, both for the photo-z
results and for the true z results. The photo-z result is shown in 
Fig.~\ref{fig:omvsw}, where the MICE cosmology is inside the 
1-$\sigma$ contours. These contours have been obtained fitting the 
cosmology to the results presented in Fig.~\ref{fig:thbao_vs_z} (top), with 
free parameters $\Omega_M$ and $w$, and fixing all the other parameters 
to their MICE values. 

In order to perform this fit, we need to take into account the correlation
between bins induced by photo-z. To do this, we compute the migration matrix
$r_{ij}$, given by the probability that a given galaxy at bin $i$ is measured
at bin $j$ due to the photo-z error. Once the migration matrix is computed 
(just counting the number of galaxies that migrate from bin $i$ to bin 
$j$ in the simulation), the calculation of the covariance matrix $C_{ij}$ is 
straightforward. Since

\begin{equation}
N_{i}^{O}=\sum_{j=1}^{N_{bins}} r_{ij} N_{j}^{T}
\end{equation}

\noindent
where $N_{i}^{O}$ is the number of observed galaxies in bin 
$i$, $r_{ij}$ is the migration matrix and $N_{j}^{T}$ is the true number 
of galaxies in bin $j$, then 
the observed and true correlation functions are related 
by \citep{Benjamin:2010rp}:

\begin{equation}
\label{eq:wobsvswtrue}
w_{i}^{O}(\theta) = \sum_{k=1}^{N_{bins}} r_{ik}^2 
             \frac{(N_{k}^{T})^2}{(N_{i}^{O})^2} w_{k}^{T}(\theta)
\end{equation}

\noindent
where $w_{i}^{O}(\theta)$ is the observed correlation function in bin $i$
at scale $\theta$ and $w_{i}^{T}(\theta)$ is the true correlation function
in bin $i$ at scale $\theta$, given by Eq.~\ref{eq:xir2wtheta}. Then, the 
covariance for the correlation functions can be written

\begin{equation}
\label{eq:cov}
C_{ij} = \langle \Delta w^O_i(\theta) \Delta w^O_j(\theta^{\prime}) \rangle = 
\sum_{k=1}^{N_{bins}} (r_{ik}^2 r_{jk}^2) 
\frac{(N_{k}^{T})^4}{(N_{i}^{O})^2 (N_{j}^{O})^2} 
{\rm Cov}_{\theta \theta^{\prime}}
\end{equation}

\noindent
where ${\rm Cov}_{\theta \theta^{\prime}}$ is given in 
Eq.~\ref{eq:CovW}. This expression neglects the intrinsic correlations 
between bins. Whether this is a good approximation depends mainly on the 
bin widths. In our case, adjacent bin centers are separated by more than 
200 Mpc/$h$, and we find that the intrinsic cross-correlations can be 
safely ignored. Thus, with photo-z errors, the observed correlation 
functions and covariances are given by Eqs. ~\ref{eq:wobsvswtrue} and
~\ref{eq:cov}, respectively, rather than Eqs.~\ref{eq:xir2wtheta} and
~\ref{eq:CovW}.

\begin{figure}
\centering
\includegraphics[width=0.50\textwidth]{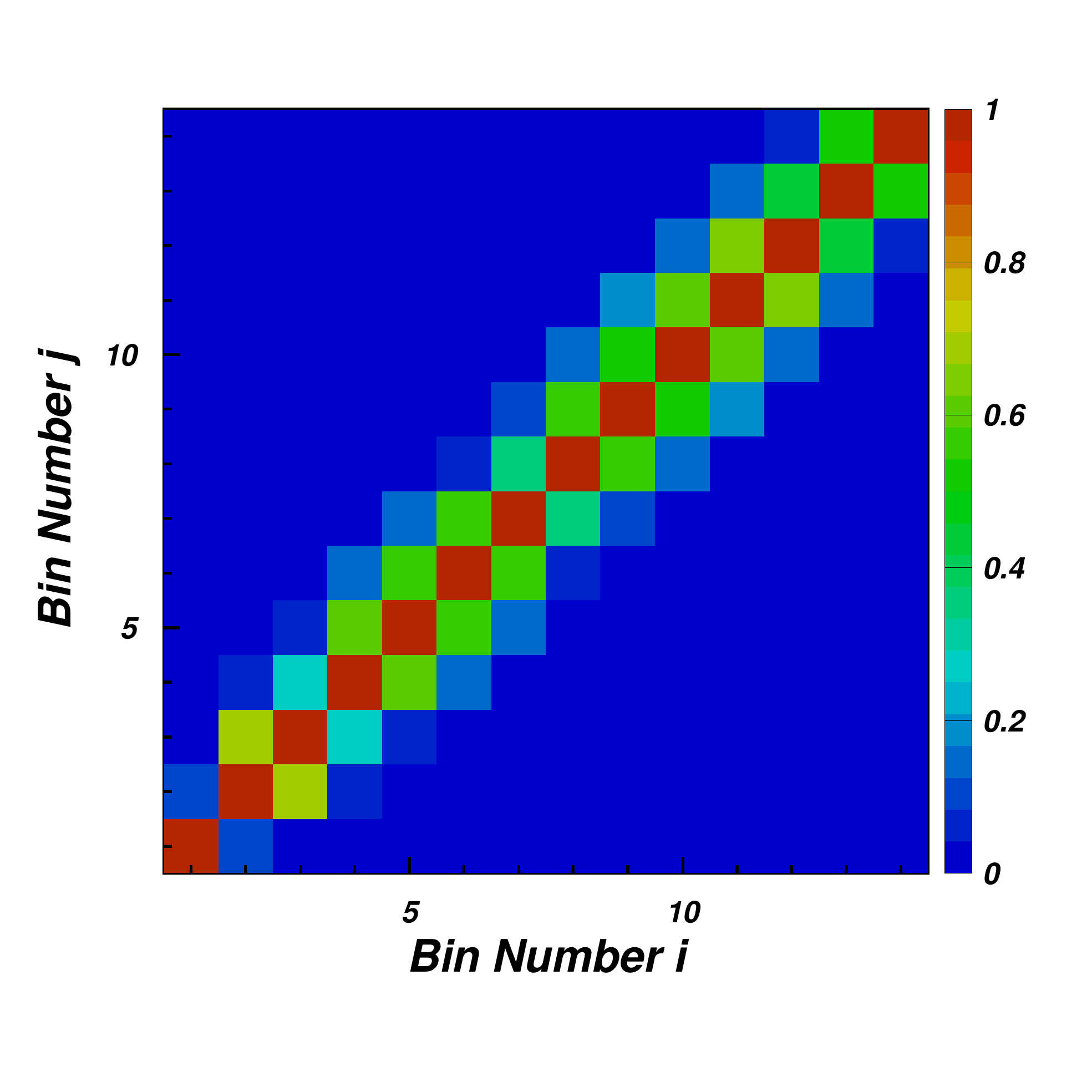}
\vspace*{-1.0truecm}
\caption{Correlation matrix for redshift bins. The correlation extends
         to 3 bins, and is taken into account in the fit of the cosmological
         parameters.\label{fig:correlation_matrix}}
\end{figure}

The resulting correlation coefficients, computed
from $C_{ij}$, for redshift bins are depicted in 
Figure~\ref{fig:correlation_matrix}. The induced correlation extends 
to 3 bins. This correlation matrix is converted to the covariance of
$\theta_{BAO}$ angles, $\widehat{C}_{ij}$, including the corresponding errors.

Therefore, we use the $\widehat{C}_{ij}$ to compute the $\chi^2$ for the 
fit in the usual way:

\begin{equation} 
\chi^2 (\Omega_M,w)  =\sum_{i,j}^{N_{bins}} 
(\widehat{\theta}^{i}_{BAO} - \theta^{i}_{BAO}) \widehat{C}^{-1}_{ij}
(\widehat{\theta}^{j}_{BAO} - \theta^{j}_{BAO})
\end{equation}

\noindent
where $\theta^{i}_{BAO}$ is the angular scale corresponding to the 
sound horizon for cosmological parameters $\Omega_M,w$ at the redshift of bin
$i$, and $\widehat{\theta}^{i}_{BAO} = \alpha \, \, \theta_{FIT}$ is the measured 
angle in bin $i$. All the results include the systematic errors in the measurements 
of $\widehat{\theta}_{BAO}$, described more in detail in the next section. The 
statistical error is correlated using the computed covariance matrix. Some of 
the systematic errors (coming from photo-z and redshift space distortions) are
also correlated, while the others are uncorrelated and are just added in 
cuadrature to the diagonal of the covariance matrix. All these numbers match 
the DES requirements. The total area of the survey, together with 
the magnitude limits, which fix the galaxy density, contribute to the 
errors on $\omega(\theta)$. Moreover, the photo-z requirements fix the 
level of correlation among redshift bins.

If we restrict to a one-dimensional analysis, and fix $\Omega_M=0.25$, then
we obtain for the equation of state of the dark energy $w=-1.05 \pm 0.14$. The
precision on this result depends on the photo-z error, which is the 
dominant source of 
systematic uncertainties. If the photo-z error is decreased, the precision in 
$w$ correspondingly improves. For a more optimistic estimation of 
$\Delta z \sim 0.03 (1+z)$ it becomes $\Delta w \sim 0.10$.

  \subsection{Systematic Errors}
  \label{subsec:sys}
We have identified several sources of systematic errors for this
methodology: 

\begin{itemize}
\item Uncertainties in the fitting procedure (method error).
\item Residual errors due to the correction in $\theta_{FIT}$ coming
      from a possible residual dependence on the cosmological 
      model (cosmology independence).
\item Uncertainties in the theory coming from the implementation
      of non-linearities (modeling error).
\item Effects of the redshift space distortions on the BAO 
      scale (z-distortions error).
\item Uncertainty in the redshift determination (photo-z error).
\end{itemize}

A brief quantitative summary of these effects is given in
Table~\ref{tab:sys}, that follows from the different analysis
discussed below.

\begin{table}
\centering
\begin{tabular}{ccc}
\hline
Systematic error & $\Delta \theta_{BAO}$ & Correlated between bins\\
\hline\hline
Parametrization            & 1.0\% & No  \\
Photometric redshift       & 5.0\% & Yes \\
Redshift space distortions & 1.0\% & Yes \\
Theory                     & 1.0\% & No  \\
Projection effect          & 1.0\% & No  \\
\hline
\end{tabular}
\caption{Estimation of various systematic errors. Some of them are
correlated between redshift bins (photo-z and z-distorions errors). The 
others are not.\label{tab:sys}}
\end{table}

     \subsubsection{Method Error}
     \label{subsub:sys_method}
To compute the systematic error associated to the parametrization method, we 
have done some further analysis on theoretical angular correlation
functions with 
the same bin widths and central redshifts as those used in the analysis
of the MICE simulation. The error associated to the method comes from 
the possible influence in the 
obtained $\theta_{FIT}$ of the range of angles used to perform the fit. To 
evaluate the error, we have varied this range for all 
14 z-bins we have, and performed the fit for each range. 

\begin{figure}
\includegraphics[width=0.50\textwidth]{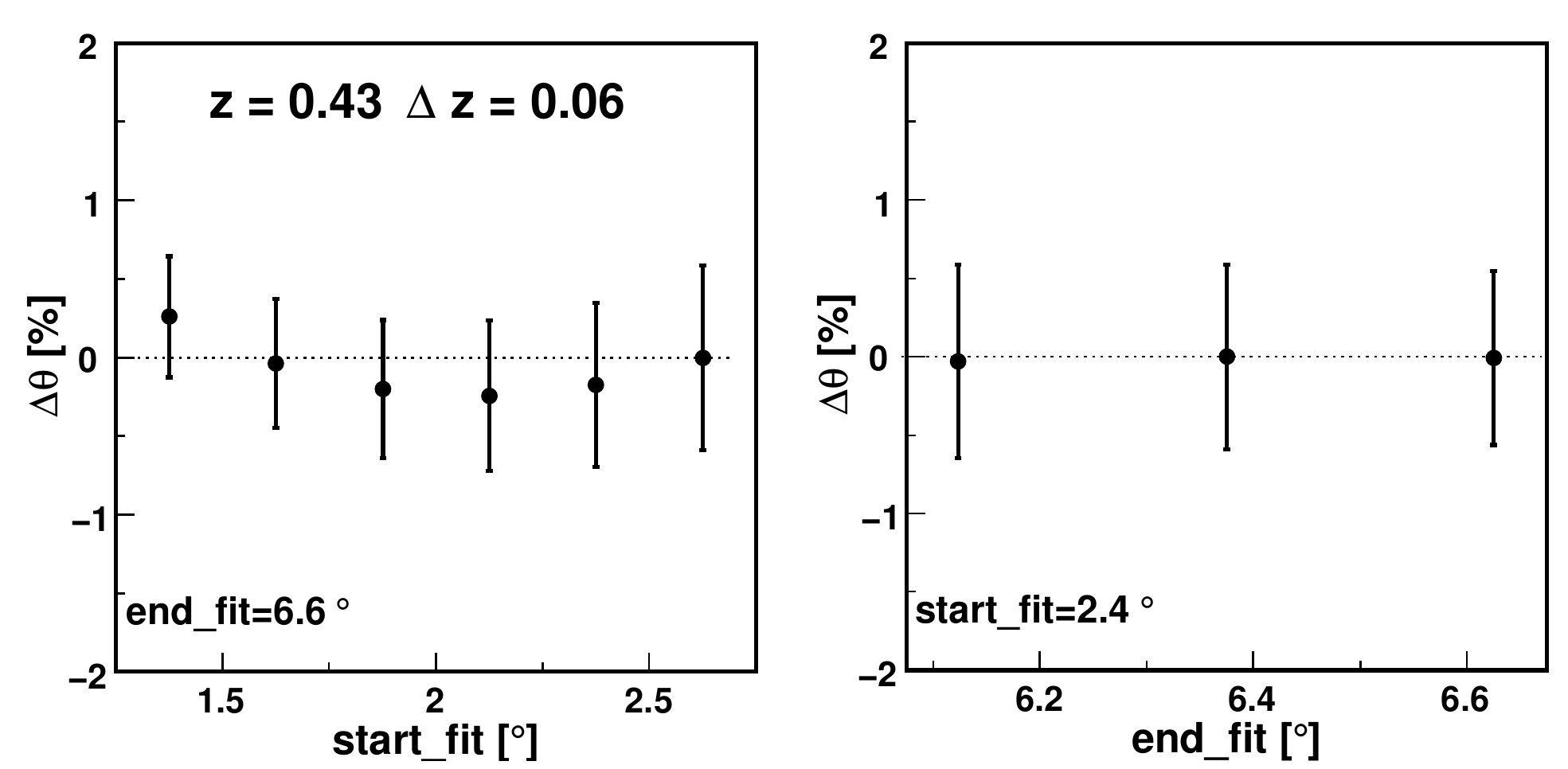}  
\includegraphics[width=0.50\textwidth]{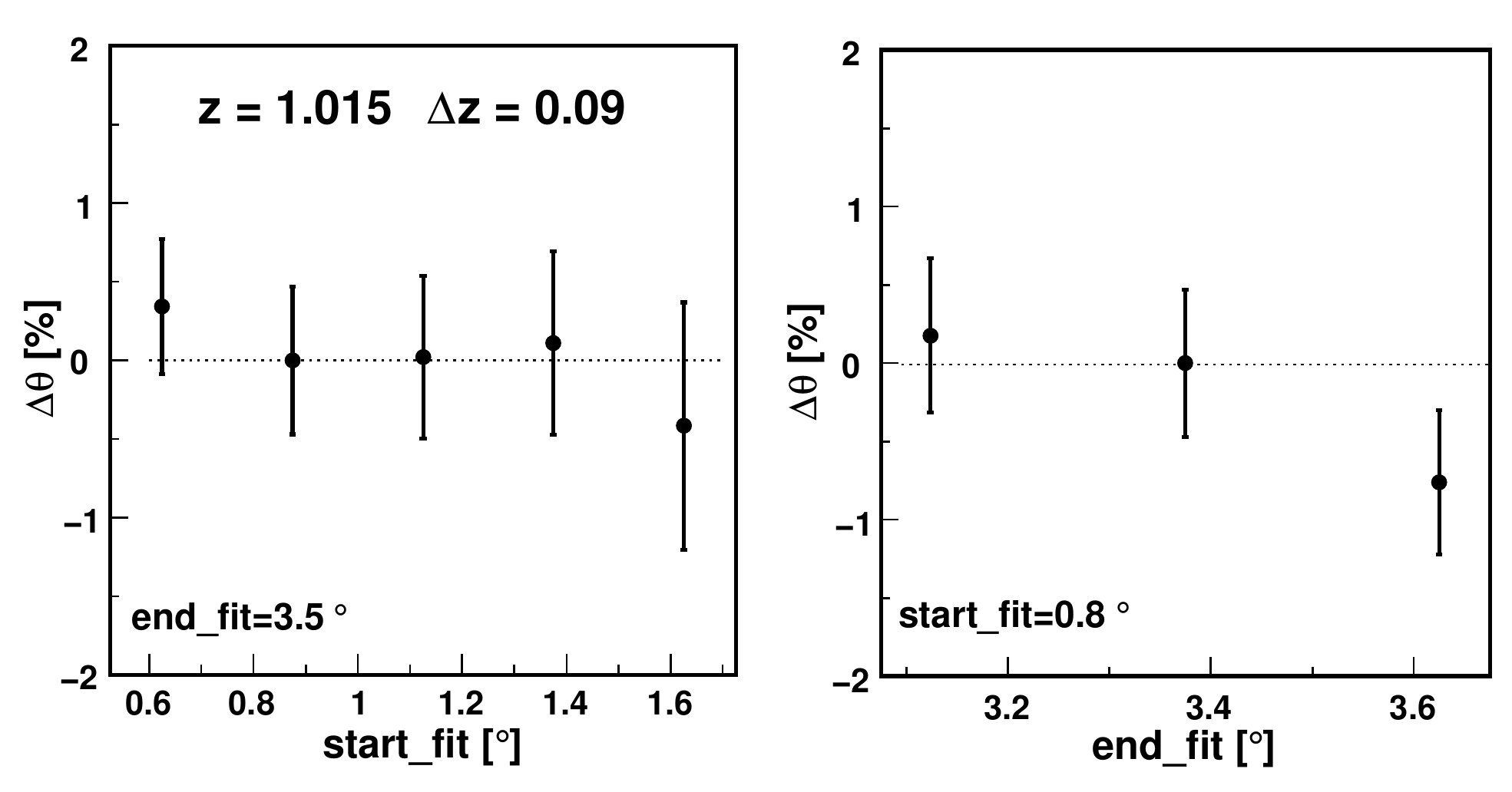}
\caption{$\theta_{BAO}$ dispersion as a function of the starting 
         and end point of the fitted region. Only two bins are 
         shown, where the maximum  of the effect is observed, at 
         redshifts 0.43 (top) and 1.015 (bottom). All the bins behave 
         in a very similar way, but with smaller variations. The 
         observed variation in the
         results is slightly smaller than 1\%, and a conservative 
         1\% systematic error is associated to the fit 
         method. \label{fig:method1}}
\end{figure}

In the decision of the range to be fitted, we have to choose a starting 
point at angles smaller than the BAO peak and an end point after the 
peak. By varying this two angles we can study how much the result vary 
with this decision. Results can be seen in Fig.~\ref{fig:method1}, where
the obtained $\theta_{FIT}$ is shown for different starting points and 
end points  of the fit, for low and high redshift. These results show
the bins where the maximum differences are observed. In all cases, the 
uncertainty is always of the order of, but slightly below, 1\%. We 
have conservatively assigned a 1\% systematic error associated to the 
method, since this error is subdominant.

     \subsubsection{Cosmology Independence}
     \label{subsub:sys_proj}
The dispersion of all the cosmological models around a common value
for the projection effect correction is small, $\leq 1\%$, as
can be seen in Figure~\ref{fig:corr_indep_cosmo}. This
dispersion propagates directly to the corrected 
$\theta_{FIT}$ as a source of uncertainty. Conservatively, we have 
asigned a systematic error of 1\% associated to 
the independence of the cosmology.

     \subsubsection{Modelling Error}
     \label{subsub:sys_modelling}
Also the error due to the uncertainty in the theory (non-linearities at 
the scale of the BAO peak) has been computed obtaining a global error 
of 1\%, estimated in a conservative way as the difference between the 
$\theta_{FIT}$ measured using linear and non-linear $\omega(\theta)$, for
the same redshift bins of the analysis. The difference for infinitesimal
bins is presented in Fig.~\ref{fig:bao_lin_nonlin}.

     \subsubsection{Photo-z Error}
     \label{subsub:photoz}
\begin{figure}
\centering
\includegraphics[width=0.50\textwidth]{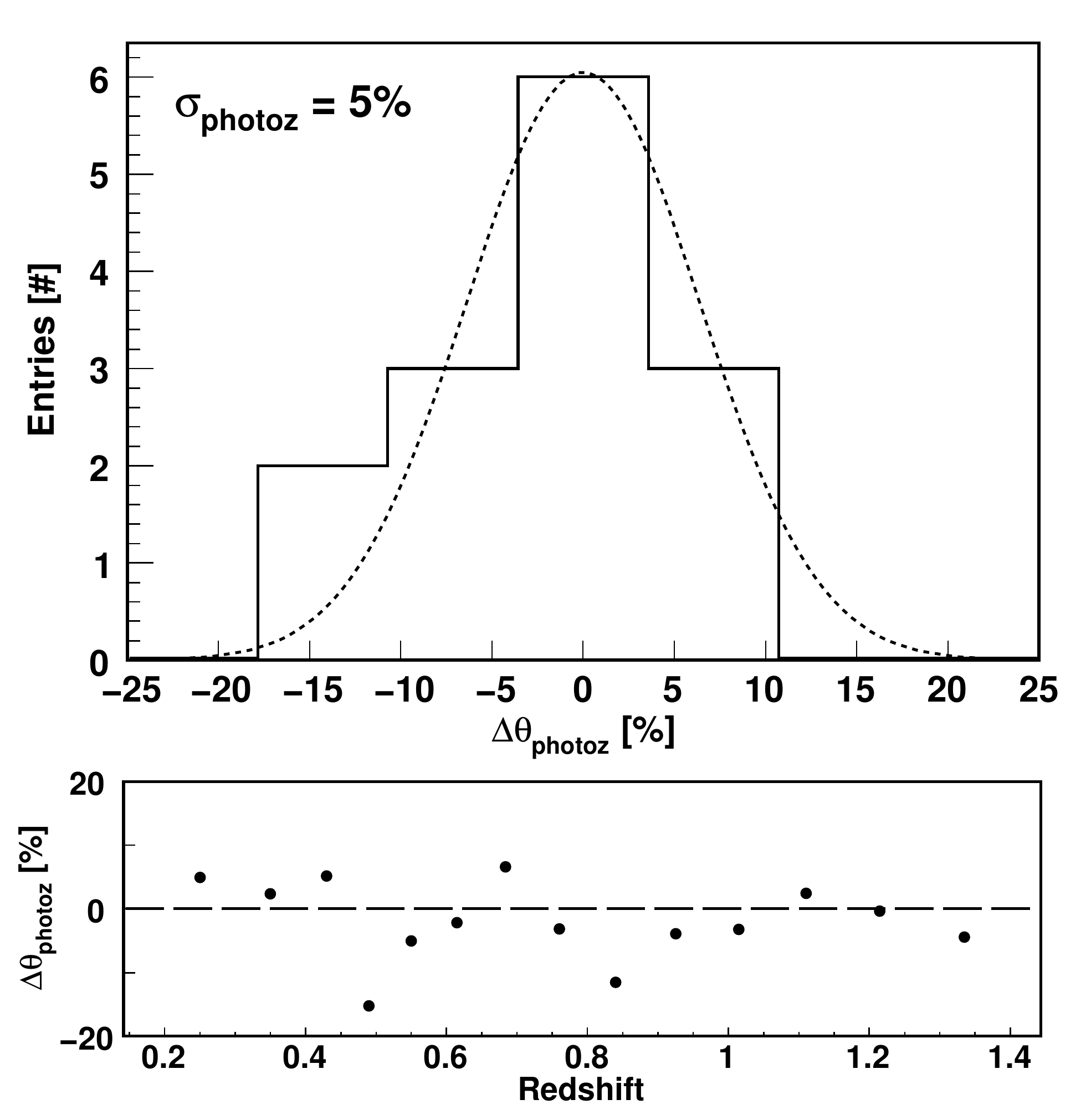}
\caption{Cumulated Difference between the measured $\theta_{FIT}$ in the
         photo-z catalog and the true z catalog for the 14 bins
         of the analysis (top). The dispersion is 5\%, that we use as 
         the systematic error. This error is taken as constant with
         redshift, since there is no clear dependence when the differences
         are plotted as a function of redshift 
         (bottom).\label{fig:photoz_dispersion}}
\end{figure}

To compute the systematic error associated to photo-z measurement, we
have analysed the simulation using the true redshifts in the same
way we do for the photo-z. We can estimate the deviation in
$\theta_{FIT}$ due to photo-z errors computing the difference between
the result obtained using the catalog with true-z and the result
obtained using the catalog with photo-z. In 
Fig.~\ref{fig:photoz_dispersion} (top) we can see the distribution of those
differences. Photo-z uncertainties do not introduce a significant 
bias, but they do introduce a dispersion around the 
central value. This 
dispersion is around 5\%, which we assign as  
the systematic error coming from the photometric error. We are assuming
here that this error is redshift independent, which given the scope of the
current analysis is reasonable, as shown in 
Fig.~\ref{fig:photoz_dispersion} (bottom). The differences between the
results of the analysis using photo-z and true z are shown as a function of
the redshift, and there is no indication of any redshift dependence. This
estimation could be further refined
by using many simulations and repeating this study for each redshift
bin as many times as simulations we use. This is out of the scope of 
this work, where we can only indicate the order of magnitude of the
systematic errors, but should be done for a real survey, since this
will be the dominant source of systematic uncertainty.

     \subsubsection{Other Systematic Errors}
     \label{subsub:other}
There are some other potential systematic errors, like those arising from
the uncertainty in the description of the redshift space distortions, the 
gravitational lensing magnification or those mainly associated to the used 
galaxy sample, as the bias of the population. 

The uncertainties in the description of the redshift space distortions
can induce a systematic error in the determination of the sound horizon
scale. To evaluate it, we have changed the parameters of the z-distortions
within their uncertainties, finding a value of 
$\Delta \theta_{BAO} \sim 0.5-1 \%$, in good agrreement with the effect seen
in Fig.~\ref{fig:z_dist}.

At the BAO scales it is a good approximation to consider that the bias 
is scale independent~\citep{2009MNRAS.392..682C,crocce10}, and given the 
width of 
the redshift bins, we can safely use a redshift independent bias inside 
each one. The redshift dependence of bias may appear in a different 
normalization of the correlation function between different redshift 
bins. Since we do not care about the overall normalization and only about 
the position of the peak, we are not sensitive to $b(z)$. We have tested
that the change in overal amplitude of the correlation function does not
affect to our determination of the sound horizon scale, even if the error
in the correlation function is changed. Moreover, we have
tested the effect of a scale dependent bias, introducing artificially
the effect in the correlation functions, using an approximated Q-model
with the determination of parameters of~\citet{2009MNRAS.392..682C}. The 
bias variation with $\theta$ in the fit region of $\omega(\theta)$ ranges 
from 1\% to 6\%, and induces a change in the determination of the sound 
horizon scale below 1\%, well within statistical uncertainties. 

Finally, the gravitational lensing magnification can affect 
the $\omega(\theta)$~\citep{2008PhRvD..77b3512L} introducing
correlations between bins at different redshifts. However, the effect
of the magnification is concentrated at very low angles, and the magnitude
of the induced correlation is much smaller than the correlation
induced by photo-z, which has already been studied and used in this
analysis. Therefore, the position of the BAO peak is free of systematics 
coming from magnification. Consequently, all this effect is not 
considered here.

\section{Conclusions}
\label{sec:conclusions}    
We have developed a new method to measure the BAO scale in the angular
two-point correlation function. This method is adapted to photometric
redshift surveys, where the information along the line of sight is lost
due to the photo-z precision, and only the angular information 
survives, although it can be applied to any survey. Two main results
are found. First, the sound horizon scale can be recovered from the
non-linear angular correlation functions to a precision $\leq$0.75\% 
applying the parametric fit described in the text, for any cosmological 
model for infinitesimal redshift shells ({\it i.e.} for the 3D 
correlation function). This is not totally surprising. It can be 
understood if we approximate the 3D linear correlation as
a sum of a BAO gaussian and a baseline that is well described by a 
power-law. If we neglect mode coupling effects, the non-linear correlation
is given by a convolution with a gaussian damping, which leave unchanged
the Gaussian position. Under this assumption our parametric fit 
should be able to recover the true BAO positions without any 
bias. Second, the shift of the BAO 
peak due to projection effects has a universal shape, does not depend
on the cosmological model, and only depends 
on the redshift and the redshift bin width. This can be used to 
correct the result 
obtained for wide photo-z bins and recover the true sound horizon
scale. The method has been tested with a mock catalog built upon a large 
N-body simulation provided by the MICE collaboration, with characteristics
similar to those expected in DES. The true cosmology is recovered within 
1-$\sigma$. The correlation between redshift bins and a preliminary 
evaluation of the systematic errors have been included in this 
study, and we find that the most important systematic error arises 
from the photo-z precision. The method is very promising and very 
robust against systematic uncertainties. 

Note that our analysis over the MICE simulations was in real, rather
than in redshift space. Over a comoving output, rather than over
an evolving distribution. Over dark matter particles, instead of 
galaxies. We believe that all this simplifications are not essential 
limitations to the method presented here, as we have shown that
both the modeling and the error analysis are quite generic and
work in real, redshift and photo-z space. If anything, redshift space 
distortions (and possibly biasing) produce larger amplitudes, in 
which case our results are conservative in the sense that more realistic 
simulations (or observations) will produce smaller error bars than the 
ones presented here.

\section*{Acknowledgements}
\label{sec:acknowledgements}  
We thank the Large Scale Structure Working Group of the DES Collaboration
for the support, help and useful discussions on several stages of this analysis.

We thank the Spanish Ministry of Science and Innovation (MICINN) for
funding support through grants AYA2009-13936-C06-01, 
AYA2009-13936-C06-03, AYA2009-13936-C06-04, AYA2009-13936-C06-06 and 
through the Consolider Ingenio-2010 program, under project CSD2007-00060. We 
acknowledge the use of data from the MICE simulations, publicly 
available at http://www.ice.cat/mice. JGB thanks the Department of 
Theoretical Physics at Universit\'e de Gen\`eve for their generous 
hospitality during his sabbatical year. FdS would like to thank the 
DES-Brazil collaboration for useful discussion and support during this 
work. FdS is supported by the Brazilian National Research Council (CNPq).


\end{document}